\documentclass[aip,reprint]{revtex4-1}

\usepackage[utf8]{inputenc}
\usepackage{amsmath}
\usepackage{amsfonts}
\usepackage{amssymb}
\usepackage{float}
\usepackage{graphicx}
\usepackage{color}
\usepackage{epsfig}
\usepackage{hyperref}

\newcommand{\be}{\begin{equation}}
\newcommand{\ee}{\end{equation}}
\newcommand{\bea}{\begin{eqnarray}}
\newcommand{\eea}{\end{eqnarray}}
\newcommand{\ba}{\begin{eqnarray*}}
\newcommand{\ea}{\end{eqnarray*}}

\newcommand{\eps}{\varepsilon} 

\newcommand{\aver}[1]{\langle {#1} \rangle}

\newcommand{\es}[1]{\begin{split}#1\end{split}}
\newcommand{\beq}{\begin{equation}}
\newcommand{\eeq}{\end{equation}}

\newcommand{\lp}{\left(}
\newcommand{\rp}{\right)}
\newcommand{\lsq}{\left[}
\newcommand{\rsq}{\right]}
\newcommand{\lbr}{\left\lbrace}
\newcommand{\rbr}{\right\rbrace}
\newcommand{\da}{\dagger}
\newcommand{\bma}{\begin{pmatrix}}
\newcommand{\ema}{\end{pmatrix}}

\newcommand{\bra}[1]{\langle #1 |}
\newcommand{\ket}[1]{| #1 \rangle}

\newcommand{\rw}{\rightarrow}
\newcommand{\oh}{\frac{1}{2}}

\newcommand{\pt}{\partial _t}
\newcommand{\dt}{ dt \,}
\newcommand{\dtp}{ dt' \,}

\newcommand{\tr}{\text{tr}}

\usepackage{bbold}
\newcommand{\id}{\mathbb{1}}

\newcommand{\sign}{\text{sign}}

\newcommand{\nm}{{\bar{M}}}

\newcommand{\hc}{\text{hc}}

\newcommand{\mcU}{\mathcal{U}}
\newcommand{\mcV}{\mathcal{V}}
\newcommand{\mcD}{d}
\newcommand{\lind}{\mathcal{L}}
\newcommand{\dbar}[1]{\overline{\overline{{#1}}}}

\newcommand{\cmm}[1]{\textcolor{black}{#1}}

\newcommand{\ig}[4][]{
\begin{figure}[H]
\centering
\includegraphics[width=#3 \linewidth]{#2}
\caption{#4 \label{#1}}
\end{figure}
}

\draft 

\begin{document}


\title{Quantum Impurity Models coupled to Markovian and Non Markovian Baths}



\author{Marco Schiro}\thanks{Email: marco.schiro@ipht.fr\\ On Leave from: Institut de Physique Th\'{e}orique, Universit\'{e} Paris Saclay, CNRS, CEA, F-91191 Gif-sur-Yvette, France}
\affiliation{JEIP,  USR  3573  CNRS,  Coll\'{e}ge  de  France,  PSL  Research  University,11,  place  Marcelin  Berthelot,  7
5231  Paris  Cedex  05,  France}
\author{Orazio Scarlatella}\thanks{Email: orazio.scarlatella@ipht.fr}
\affiliation{Institut de Physique Th\'{e}orique, Universit\'{e} Paris Saclay, CNRS, CEA, F-91191 Gif-sur-Yvette, France}



\date{\today}

\begin{abstract}

We develop a method to study quantum impurity models, small interacting quantum systems \cmm{bilinearly} coupled to an environment, in presence of an additional Markovian quantum bath, with a generic non-linear coupling to the impurity. 
We aim at computing the evolution operator of the reduced density matrix of the impurity, obtained after tracing out all the environmental degrees of freedom. First, we derive an exact real-time hybridization expansion for this quantity, which generalizes the result obtained in absence of the additional Markovian dissipation, and which could be amenable to stochastic sampling through diagrammatic Monte Carlo. Then, we obtain a Dyson equation for this quantity and we evaluate its self-energy 
with a resummation technique known as the Non-Crossing Approximation.
We apply this novel approach to a simple fermionic impurity coupled to a zero temperature fermionic bath and in presence of Markovian pump, losses and dephasing.
%
\end{abstract}


\maketitle 



%
%

%

\section{Introduction}

Small interacting quantum systems coupled to external environments represent basic paradigms of transport, dissipation and non equilibrium phenomena. Understanding the dynamical behavior of these open quantum systems is therefore crucial in many different physical contexts where the idealization of an isolated quantum system obeying perfectly unitary quantum dynamics is either to restrictive or unable to capture the fundamental physics.

In condensed matter physics the motivation comes from studying models of quantum dissipation and macroscopic quantum tunneling in the early days of Caldeira-Leggett and spin-boson models~\cite{Caldeira_Leggett_PRL81,Leggett-RMP} which keep attracting a lot of interest~\cite{LEHUR2018451} or from diluted magnetic impurities in metals~\cite{Hewson_book} and transport through quantum dots and single molecules attached to leads~\cite{Goldhaber_Gordon_nature98, GoldhaberGordon_prl08_noneq,Florens_C60,IftikharEtAlScience2018} leading to fermionic realizations of so called quantum impurity models.  These consist of a small quantum systems 
with few interacting degrees of freedom, the impurity,  coupled via hybridization to a gapless reservoir of fermionic or bosonic excitations. The dynamical correlations of such reservoirs, which decay in time as a power law at zero temperature and feature strong memory effects~\cite{CohenRabaniPRB11}, together with a local many body interaction, make quantum impurity physics highly non-trivial. Nevertheless, methods to solve the dynamics of quantum impurity models have flourished in recent years, mainly driven by the developments of Diagrammatic Monte Carlo~\cite{Gull_RMP11,muhlbacherRabaniPRL2008,Keldysh_short,Werner_Keldysh_09,CohenEtAlPR13,ProfumoEtAlPRB15,
CohenEtAlPRL15,ThetaJCP17_1,ThetaJCP17_2}.

On a different front, recent advances in quantum optics, quantum electronics and quantum information science have brought forth novel classes of driven open quantum systems in which excitations are characterized by finite lifetime due to unavoidable losses, dephasing  and  decoherence  processes  originating from their coupling to an external electromagnetic environment. Examples include atomic and optical systems such as ultracold gases in optical lattices~\cite{BlochDalibardNascimbeneNatPhys12} or trapped ions~\cite{BlattRoosNatPhys12}, as well as solid state systems such as arrays of nonlinear superconducting microwave cavities~\cite{AndrewNatPhys,LeHurReview16}.
In these settings, the dissipative processes associated to the external environment can be very well described in terms of a Lindblad master equation for the system density matrix~\cite{BreuerPetruccione}. A major effort here is to conceive situations in which coupling to a quantum environment can involve non-linear combinations of system operators thus mediating effective interactions which act as a resource for quantum state preparation and to engineer a desired steady state. Such a dissipation engineering is actively investigated in quantum optics~\cite{DiehlEtalNatPhys08,VerstraeteWolfCiracNatPhys09,Siddiqi_quantum_bath_engineering}. This has stimulated a new wave of interest around open Markovian quantum systems at the interface between quantum optics and condensed matter physics.


The examples above represent two well studied, yet substantially separated, paradigms of open quantum systems. At the same time much less is known about the interface between those two settings, namely the interplay between Markovian dissipation and the coupling to a fully structured, frequency dependent, Non-Markovian quantum bath, especially for what concerns the emergent many body physics.  Interest around this new kind of quantum impurity problems has recently grown~\cite{NakagawaEtAlPRL18,TonielliEtAlPRL19,HeinrichEtAlPRL19}. Such a question is potentially relevant in a number of contexts. From one side, mesoscopic quantum devices have been successfully coupled to electromagnetic resonators hosting dissipative photon fields
~\cite{Delbecq_prl11,SchiroLeHurPRB14,BruhatEtAlPRX16,CottetEtAlReview17} offering the possibility to investigate the fate exotic many body phases such as the Kondo effect in presence of Markovian dissipation. 

On the other hand, in the context of quantum optics and quantum information the role of Non-Markovian bath correlations has been recently attracting enormous interest~\cite{LiuEtAlNatPhys11,RivasHuelgaPlenioRepProgPhys14,BreuerEtalRMP16} and there is urgent need to develop novel theoretical approaches to address this question.

\begin{figure}[t]
\begin{center}
\epsfig{figure=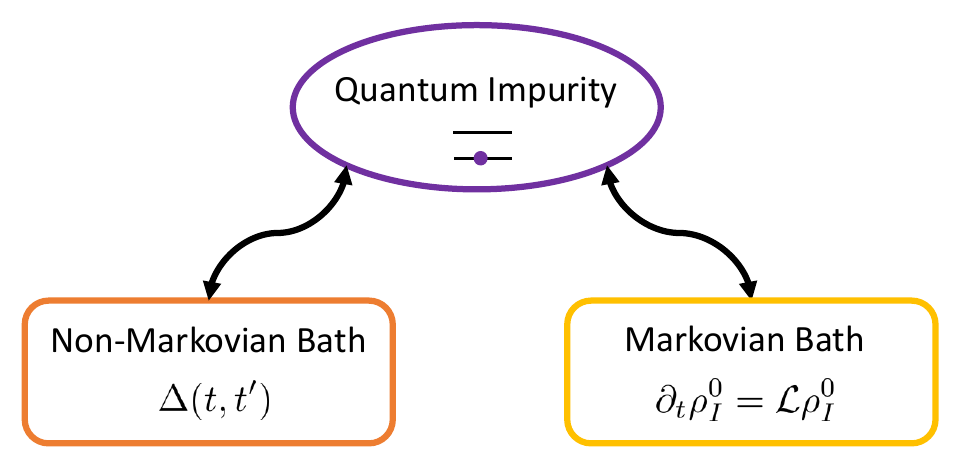,scale=0.5}
\caption{Schematic plot of the setup considered in this manuscript. A small quantum system (impurity) is (i) \cmm{bilinearly} coupled to a quantum bath whose non-trivial correlations, encoded in the hybridization function $\Delta(t,t')$, lead to Non-Markovian behavior  and (ii) coupled non-linearly to a Markovian bath whose effect on the impurity is described by a Lindblad master equation. The resulting quantum impurity model with mixed Markovian and Non-Markovian dissipation is studied using hybridization expansion techniques.}
\label{fig:image}
\end{center}
\end{figure}

With these motivations, in this work we focus on a model for a quantum impurity coupled to two kind of external environments, as depicted in figure \ref{fig:image}: a quantum bath described by a set of non interacting modes \cmm{bilinearly} coupled to our impurity degree of freedom, as in conventional quantum impurity models, resulting in a frequency dependent Non-Markovian evolution for the impurity and a second environment, where the non interacting bath modes can be also non-linearly coupled to the impurity, but for which the reduced impurity dynamics would be well described by a Markovian master equation. The resulting impurity problem encodes therefore the interplay between those two kind of dissipative mechanisms and the local interaction on the impurity.

Taking inspiration from recent developments in quantum impurity physics~\cite{Gull_RMP11,muhlbacherRabaniPRL2008,schiroFabrizioPRB2009,Werner_Keldysh_09} we develop here an hybridization expansion for the real-time evolution operator of the impurity, obtained after tracing out both the non-Markovian environment, as usually done in the literature, as well as the Markovian bath. The final result naturally generalizes the well known real-time hybridization expansions to this mixed Markovian/Non-Markovian context. In addition to its own interest and potential for the development of diagrammatic Monte Carlo sampling, this expansion allows us to formulate a self-consistent resummation technique for the real-time impurity evolution operator based on the Non-Crossing Approximation used in the context of quantum impurity models~\cite{Bickers1987,Nordlander1999,Eckstein_NCA_PRB10,rueggMillisPRB2013,strandWernerPRX2015,peronaciPRL2018}. We derive and discuss in details this approach and test it on a simple fermionic model coupled to a zero temperature bath and in presence of Markovian dissipation.

The paper is organized as follows. In Section II we introduce the model and the general formulation of the problem. Section III is devoted to derive the hybridzation expansion for the real-time impurity propagator, after tracing out the non-Markovian bath (section III.A) and the Markovian environment (Section III.B). In Section IV we develop a self-consistent resummation based on the Non-Crossing Approximation, while in Section V we apply this method to a simple model.

\section{Model and General Formulation of the Problem}

We consider a model of a quantum impurity, a small quantum system with a finite number of bosonic/fermionic degrees of freedom $[d_{a},d^{\dagger}_b]_{\pm}=\delta_{ab}$ and with Hamiltonian $H_I[d_a,d^{\dagger}_a]$, coupled to two different quantum baths (see figure \ref{fig:image}). We will denote the Hamiltonian of the baths with $H_M$ and  $H_{\bar{M}}$, where the subscripts refer to the fact that $\nm$ is a non-Markovian bath and $M$ is a Markovian one. We describe the two environments as a collection of non-interacting bosonic/fermionic modes, (respectively if the impurity is bosonic/fermionic)
\begin{align}
\label{eq:bathCoupling}
 H_M &= \sum_p \omega_p b_p^\da b_p & 
 H_\nm &= \sum_k \eps_k c_k^\da c_k 
\end{align}
The total Hamiltonian therefore reads 
$$
H = H_I + H_M+ H_{IM} + H_{\bar{M}}  + H_{I\bar{M}}
$$
 where we have introduced the two coupling terms between the impurity and the $M$ and $\bar{M}$ baths. 
We will consider the impurity to be \cmm{bilinearly} coupled to the $\bar{M}$ bath, i.e. through a coupling Hamiltonian of the form
\begin{equation}
H_{I\bar{M}} = \sum_{ka}  V_{ka} \left(d_a^\da  c_{k} + \hc \right)
\end{equation}
\cmm{while the coupling between the impurity and the $M$ bath is taken of the most general form for which one can derive a Lindblad master-equation \cite{breuerPetruccione2010}: 
\begin{equation}
\label{eq:markCoupling}
H_{IM} = \sum_{\alpha} X_\alpha B_\alpha
\end{equation}
with $X_\alpha = X_\alpha^\da$, $B_\alpha = B_\alpha^\da$ generic operators respectively of the impurity and of the Markovian bath.}




Defining the time evolution operator of the entire system as $U(t,0)= e^{- i H t}$ and given an initial condition for the system density matrix $\rho(0)$ we can formally write down the reduced density matrix of the impurity at time $t$, tracing out the degrees of freedom of the two environments
\beq \label{eq:2traces}
\rho_I(t)=\tr_{M\nm} \lsq U (t,0) \rho(0) U^\da (t, 0)  \rsq
\eeq 
from which the dynamics of simple impurity observables can be readily obtained as $O_I(t)=\tr\lsq \rho_I(t)O_I\rsq$. 
With the assumption that the initial density operator of the environment and the impurity factorizes \cite{breuerPetruccione2010}, we can define the evolution operator of the reduced dynamics
\beq 
\label{eq:reducedEvolution}
\rho_I(t)= \mcV (t,0) \rho_I(0)
\eeq
This reduced density operator and its evolution operator are the key quantities over which we will focus our attention throughout the manuscript.  Performing the trace over the environment degrees of freedom is a highly non-trivial problem. In the following we will obtain two main results.
The first one is a formal series from $\mcV$ to all orders in the coupling with the non-Markovian bath, called hybridization expansion, and the second one is a closed equation for $\mcV$, based on a self-consistent resummation of the series.

\cmm{We stress that the order in which the trace over the two environments is taken in Eq.~(\ref{eq:2traces}) is not crucial. While in this paper we will proceed by first taking the trace over the non-Markovian environment, resulting in an hybridization expansion, and then over the Markovian one, we could have equally reversed this choice and still obtain the same final result.}

\section{Hybridization Expansion}

In this section we derive a formal hybridization expansion \eqref{eq:hybExp} for the reduced density matrix of the impurity. Such an expansion is usually derived in the context of quantum impurity models coupled to a single bath, as a starting point to develop exact Monte-Carlo sampling \cite{muhlbacherRabaniPRL2008,schiroFabrizioPRB2009} or approximated resummation techniques \cite{aokiWernerRMP2014,rueggMillisPRB2013} to solve the problem. There the formulation is typically done at the level of the partition function, i.e. tracing out also the impurity degrees of freedom, while we are interested in the reduced density matrix and the evolution operator, see Eq. (\ref{eq:reducedEvolution}), therefore we will not perform such a trace, a fact that will result in some formal difference in the approach. More importantly, here the quantum impurity is also coupled to a second Markovian environment that we will need to trace out \cmm{as well and this can be done exactly} under the assumption that the $IM$ subsystem obeys a Markovian Lindblad master equation~\cite{breuerPetruccione2010,carmichaelStatistical1999}.
%
%

\subsection{Tracing Over the Non-Markovian Bath}

We begin by performing the trace over the non-Markovian environment which is quadratic in terms of bath operators and \cmm{bilinearly} coupled to the impurity. Such a trace could be performed exactly within a path integral formulation leading to an effective Keldysh action which is non-local in time. Here we proceed instead at the operator level by noticing that the trace could be taken exactly order by order in an expansion in the the coupling between the non-Markovian environment and the impurity.

In order to derive this expansion, we write down the full Hamiltonian of the system as $H=H_0+H_{\bar{M}}+H_{I\bar{M}}$, describing respectively the impurity embedded in the Markovian bath ($H_0=H_I + H_M+ H_{IM}$), the Non-Markovian environment and its coupling to the impurity. We then move to the interaction picture with respect to the Hamiltonian $H_0+H_{\bar{M}}$. \cmm{Introducing the standard time-ordering and anti-time-ordering operators $T_t$ and $\check{T}_t$,} the density operator becomes 
 \beq \label{eq:densMatEvol}
 \es{
\rho (t) = &e^{-i \lp H_0 + H_{\bar{M}} \rp t } T_t e^{-i \int_0^t \dtp H_{I\bar{M}}(t')} \rho (0) \times \\ \times & \check{T}_t e^{i \int_0^t \dtp H_{I\bar{M}}(t')} e^{i \lp H_0 + H_{\bar{M}} \rp t } 
}
\eeq 
%
%

We will perform a simultaneous expansion in powers of $H_{I \nm}(t')$ both on the left and on the right of the initial density operator $\rho(0)$. A formal way to manage a single series expansion and to write all the operators on the left side of the density operator, is to use the formalism of the Schwinger/Keldysh double contour $C(t,0)$ \cite{schwinger1961,keldysh1964diagram} (fig. \ref{fig:contour}).
Operators on the left (right) side of $\rho(0)$ are assigned a + or - label, so that the couple $(t,\gamma) \equiv t_\gamma$ with $\gamma \in \{+, - \}$, allows to locate one operator on this double time-axis.

\ig[fig:contour]{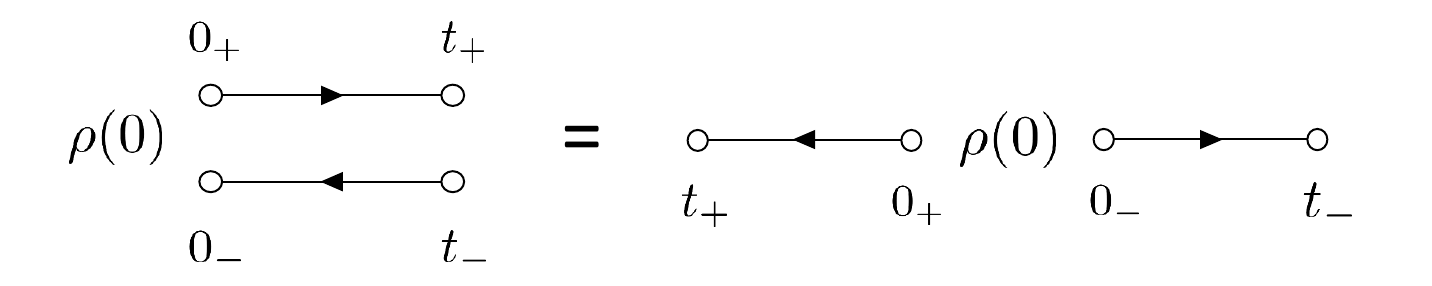}{1}{Two equivalent pictorial representations of the Schwinger/Keldysh contour $C(t,0)$, describing the non-equilibrium evolution of an initial density operator $\rho(0)$ from time $0$ to time $t$. The two branches of the contour are usually called $+$ and $-$ and they correspond to the two time evolution operators applied to the left and to the right of the initial density operator, as in eq. \eqref{eq:densMatEvol}. }


We introduce the standard Keldysh time-ordering on the contour as follows:
\beq
t_\gamma > t_{\gamma '} ' \qquad \text{if} \quad
\begin{cases}
t>t'  &\gamma = \gamma ' =+  \\ 
t<t' &\gamma = \gamma ' = -  \\ 
 \gamma = - &\gamma ' = +
\end{cases}
\eeq 
This ordering allows to define a time-ordering operator $T_C$, such that two operators, $X_1$ and $X_2$, being $X$ a creation or annihilation fermionic (bosonic) operator, anticommute (commute) under time-ordering: $T_C X_1{(t_\gamma)} X_2{(t_{\gamma '})}= \xi T_C  X_2{(t_{\gamma '})}  X_1{(t_\gamma)} $, with 
\begin{align*}
\xi &=1 &\text{for bosons} \\
\xi &=-1 &\text{for fermions}
\end{align*}
The time ordering operator is defined as follows
\beq
T_C X_1{(t_\gamma)} X_2{(t_{\gamma '})} = \begin{cases}
X_1{(t_\gamma)} X_2{(t_{\gamma '})} & \text{for $t_\gamma>t'_{\gamma'}$} \\ 
\xi X_2{(t_{\gamma '})} X_1{(t_\gamma)} & \text{for $t_\gamma<t'_{\gamma'}$} \\ 
\end{cases}
\eeq which naturally extends to the case of $n$ operators. Once time ordered, the operators belonging to the - branch of the contour have to be brought on the right side of the density matrix, as if there we were exploiting the cyclic property of a trace.

By defining contour integrals as $\int_{C(0,t)} \dt \equiv \int_0^{t_+} d t_+ - \int_0^{t_-} d t_-$, one can show that the density operator evolution \eqref{eq:densMatEvol} can be written in the compact form $$\rho(t) = T_C e^{-i \lp H_0 + H_{\bar{M}} \rp (t_+ -t_-) } e^{-i \int_{C(0,t)} \dtp H_{I\bar{M}}(t')} \rho (0)$$
%
%
and, accordingly, the evolution operator defined in \eqref{eq:reducedEvolution} can be written as 
\small
\beq \es{
\mcV(t,0)  \rho_I(0)=\tr_{M\nm} \lsq T_C   e^{-i H_0  (t_+ -t_-) }e^{-i \int_{C(0,t)} \dtp H_{I\bar{M}}(t') }\rho (0) \rsq  
}
\eeq
\normalsize
In order to perform the partial trace on the non-Markovian environment, we assume that at time $t=0$ there's no entanglement between the non-Markovian bath and the rest of the system, such that the density operator factorizes $\rho(0) = \rho_{IM}(0) \otimes \rho_\nm (0)$, with $\rho_\nm (0)$ quadratic in bosonic/fermionic operators. 
Initial thermal states could be taken into account considering a third, imaginary time branch of the non-equilibrium contour \cite{DANIELEWICZ1984305,DANIELEWICZ1984239,wagnerPRB1991}, but this is beyond the interest of this work. 

We then Taylor-expand the time-ordered exponential in power of the impurity-bath hybridization $ H_{I\bar{M}}$ and perform the trace over the bath degrees of freedom, which immediately reduces the expansion to only even order terms. Then using Wick's theorem and performing the sums over $\{b_i, b_i'\}$, we can write the final result in terms of the bath hybridization function
\cmm{\beq \label{eq:hybridizationFunction}
\Delta^{\gamma' \gamma}_{a' a} (t',t) \equiv \sum_{b'} V_{a'b'} V^*_{a b'} G_{b'} (t'_{\gamma'},t_{\gamma})
\eeq
where $G_{b'}(t'_{\gamma'},t_{\gamma})= -i \langle T_C c_{b'}(t'_{\gamma'}) c_{b'}^\da (t_\gamma) \rangle$} is the contour ordered bath Green's function. Finally, we obtain the hybridization expansion \cite{aokiWernerRMP2014,schiroFabrizioPRB2009, schiroPRB2009}:

\begin{widetext}
\beq
\label{eq:hybExp_partial}
\es{
\mcV(t,0) \rho_I  (0 ) &= \sum_{k = 0} \frac{(-i)^k}{ k !^2 } \sum_{\gamma_1 \dots \gamma_k'} \prod_i \gamma_i \gamma_i' \sum_{a_1 \dots a_k'}   \int_0^t dt_1 ... \int_0^t dt_k'  \tr_{M} \lsq  T_C e^{- i H_0 (t_+ -t_-) } d_{a_k' }^\da ({t_k'}, {\gamma_k'} ) d_{a_k }({t_k},{\gamma_k}) ...\, d_{a_1 } ({t_1},{ \gamma_1})   \rho_{IM}(0)\rsq \times  \\ &\times \sum_{\sigma \in P }  \xi^{\sign{(\sigma)}} \Delta_{a_1' a_{\sigma (1)}}^{\gamma_1' \gamma_{\sigma (1)}} (t_1 ', t_{\sigma(1)} ) \dots \Delta_{a_k' a_{\sigma (k)}}^{\gamma_k' \gamma_{\sigma (k)}} (t_k ', t_{\sigma(k)} )
}
\eeq
\end{widetext}
$\gamma_i, \gamma'_i$ are contour indices $\gamma \in \{ +,- \} $.
We notice that the hybridization function \cmm{$\Delta^{\gamma_i', \gamma_{\sigma(i)} }_{a_i' a_{\sigma(i)}}(t'_i,t_{\sigma(i)})$} connects the \cmm{ $d_{a_{\sigma(i)}} ({t_{\sigma(i)}},{\gamma_{\sigma(i)}})$ } operator with the \cmm{$d_{a_i'}^\da({t'_i},{\gamma'_i})$} one. We can interpret this construction as follows. The $d$ operator creates a "hole" in the impurity, which is propagated through the system and then annihilated by a $d$ operator. To this hole it corresponds (from the definition of $\Delta$) a particle of the environment which is created, propagated and annihilated. Thus, the series eventually describes processes in which particles hop from the impurity to the environment and back to the impurity.

\subsection{Tracing over the Markovian bath}

\subsubsection{Super-operators formalism}
It is useful to describe time-evolution using super-operators, as these are natural objects to describe the dynamics of open systems and since they provide a useful framework to work out the trace on the Markovian environment in eq. \eqref{eq:hybExp_partial}. We call super-operator an operator that acts on an operator, rather then on quantum state. The focus is shifted from the standard evolution operator $U(t,0) = e^{-i H t}$, which evolves a pure state (a ket) in time, to the super-operator $\mcU (t,0)$ which time-evolves a density operator and is defined by
\beq
\rho(t) = U(t,0) \rho(0) U(0,t) \equiv \mcU (t,0) \rho(0)
\eeq


We can write a generic time-ordered string of operators, like it appears in eq. \eqref{eq:hybExp_partial}, in the Schrödinger's picture and in a compact form, using the super-operators notation. This comes at the price of introducing some notation.

We promote $d$, $d^\da$ operators in the Schrödinger's picture to super-operators $\mcD_\gamma$, $\mcD^\da_{\gamma}$, with an contour index $\gamma$ reminiscent of the branch the original operators belonged to:
\beq 
\mcD_{\gamma} ^{\da} \lsq \bullet  \rsq = \begin{cases}
d ^{\da} \bullet & \text{if } \gamma=+  \\
\bullet  d ^{\da}  & \text{if } \gamma= -  
\end{cases}
\eeq 
We trivially generalize the contour time-ordering operator $T_C$ to the super-operators notation
\beq \es{
&T_C X_{1(t,\gamma)} \mcU_0(t,t') X_{2(t',\gamma')} =\\ &= \begin{cases}
X_{1(t,\gamma)} \mcU_0(t,t') X_{2(t',\gamma')} & \text{for $(t,\gamma)>(t',\gamma')$} \\ 
\xi X_{2(t',\gamma')} \mcU_0(t',t) X_{1(t,\gamma)} & \text{for $(t,\gamma)<(t',\gamma')$} \\ 
\end{cases} }
\eeq
The $X_{t,\gamma}$ super-operators are objects in Schrödinger's picture and their time label $t$ is just meant to know how to order them.

We also need to introduce a further "forward" time-ordering operator $T_F$, that orders two super-operators according to their time labels $t,t'$, regardless of their contour index:
\beq \es{
&T_F X_{1(t,\gamma)} \mcU_0(t,t') X_{2(t',\gamma')} = \\ &= \begin{cases}
X_{1(t,\gamma)} \mcU_0(t,t') X_{2(t',\gamma')} & \text{for $t>t'$} \\ 
X_{2(t',\gamma')} \mcU_0(t',t) X_{1(t,\gamma)} & \text{for $t<t'$} \\ 
\end{cases} }
\eeq
This definition is the same for both fermions and bosons, with no extra minus signs for fermions.

Using these definitions, we can write the following identity
\beq 
\label{eq:sopString}
\es{
&T_C e^{- i H_0 (t_+ -t_-) }  d^\da (t_k' ) \dots   d (t_1  ) \rho_{IM}(0) = \\ =  & T_F T_C \mcU _0(t,t_k') \mcD^\da_{t_k ' \gamma_k' } \mcU_0(t_k ', t_{k-1}) \dots \mcD_{t_1 \gamma_1} \mcU_0 (t_1,0) \rho_{IM}(0)
 }
\eeq

The second line is a chain of subsequent time-evolutions operated by $\mcU_0$, going overall from time $0$ to time $t$, alternated with the application of $\mcD_\gamma,\mcD^\da_\gamma$ super-operators. 
We remark that the two time-order operators $T_C$ and $T_F$ do not commute.
In order to evaluate the second line of eq. \eqref{eq:sopString}, one has first to order the super-operators according to $T_C$; this first ordering is necessary in order to compute the non-trivial sign factor obtained by swapping fermionic operators. Then the super-operators must be re-ordered according to the "forward" time-ordering operator $T_F$. This insures that, in order to evaluate eq. \eqref{eq:sopString}, one has to apply only forward in time evolution super-operators. 

\subsubsection{Performing the partial trace}

We now aim at performing the partial trace on the Markovian environment which is left in eq. \eqref{eq:hybExp_partial}.
This trace is taken on a contour time-ordered string of impurity operators. The latter are nevertheless evolved by the joint dynamics of the impurity plus the remaining bath, making the partial trace non-trivial to evaluate. Assuming that the impurity-bath dynamics is governed by a Lindblad master equation, then the partial trace becomes trivial  \cite{gardinerCollettPRA1985,carmichaelStatistical1999}. We report a proof here as this is a crucial step to obtain the hybridization expansion \eqref{eq:hybExp}.
We recall the Lindblad master equation \cite{breuerPetruccione2010,carmichaelStatistical1999}: 
\beq
\label{eq:masterEquation}
\pt \rho_I^0 = -i \lsq H_I, \rho_I^0 \rsq + \sum_\alpha \gamma_\alpha \lp  L_\alpha \rho_I^0 L^\da_\alpha - \oh \lbr L_\alpha^\da L_\alpha , \rho_I^0 \rbr \rp
\eeq
where $L_\alpha $ are the jump operators, microscopically determined by the environment-impurity coupling \eqref{eq:markCoupling}. $\rho_I^0(t)$ must not be confused with $\rho_I (t)= \tr_M \rho_{IM}(t)$, as the former is the density operator obtained by evolving $\rho(0)$ in presence of the Markovian environment alone. $\rho_I (t)$ instead is obtained by evolving $\rho(0)$ with a dynamics that includes both the Markovian and non-Markovian environments.
Defining the Markovian evolution super-operator, $\mcV_0(t-t') = e^{\lind \lp t -t'\rp } $, with $t>t'$, then 
\beq
\rho_I^{0}(t) = \mcV_0(t-t') \rho_I(0) 
 \eeq
We remark that $\mcV_0$ depends only on time differences as it satisfies \eqref{eq:masterEquation}.
This is equivalent to
\beq
\label{eq:lindDyn}
\tr_M \rho_{IM}^0 (t) = \tr_M \lsq \mcU_0(t,t') \rho_{IM}^0 (t') \rsq = \mcV_0(t-t') \tr_M \rho_{IM} ^0 (t')
\eeq
In order to show how to perform the trace of the string of super-operators in \eqref{eq:sopString}, let's assume time ordering is already enforced so that we don't have to care about it. Defining $ r_1(t) = \mcU _0(t,t_k') r_1(t_k')  $ and $r_1(t_k') = d^\da_{t_k ' \gamma_k' } \mcU_0(t_k ', t_{k-1}) \dots d_{t_1 \gamma_1} \mcU_0 (t_1,0) \rho_{IM}(0) $, we can break down the tracing operation as follows: 
\beq \es{ &\tr_M\lsq \mcU _0(t,t_k') \mcD^\da_{t_k ' \gamma_k' }  \dots \mcD_{t_1 \gamma_1} \mcU_0 (t_1,0) \rho_{IM}(0) \rsq = \\ & \tr_M r_1(t) =  \tr_M \lsq \mcU _0(t,t_k') r_1(t_k')   \rsq =   \mcV_0(t-t') \tr_M r_1(t_k')    } \eeq
The last equality is analogous to eq. \eqref{eq:lindDyn} and holds under the same assumptions leading to Lindblad master equation. 
One can iterate this procedure, as now $\tr_M r_1(t_k') = \mcD^\da_{t_k ' \gamma_k' } \tr_M \lsq \mcU_0(t_k ', t_{k-1})  r_2(t_{k-1}) \rsq$, to turn all the $\mcU_0$ super-operators in eq. \eqref{eq:hybExp_partial} in $\mcV_0$ ones. 

\subsubsection{Generalized hybridization expansion}
We then get to the final form of the hybridization expansion in presence of both a non-Markovian and Markovian environment, that is one of the main results of this work:

\begin{widetext}
\beq
\label{eq:hybExp}
\es{
 \mcV (t,0)   &= \sum_{k = 0} \frac{(-i)^k}{ k !^2 } \sum_{\gamma_1 \dots \gamma_k'} \prod_i \gamma_i \gamma_i' \sum_{a_1 \dots a_k'}   \int_0^t dt_1 \dots \int_0^t dt_k' T_F T_C  \mcV_0(t,t_k') \mcD^\da_{a_k '(t_k ' \gamma_k' )} \mcV_0(t_k ', t_{k-1}) \dots \mcD_{a_1 (t_1 \gamma_1)} \mcV_0 (t_1,0)  \times  \\ &\times \sum_{\sigma \in P }  \xi^{\sign{(\sigma)}} \Delta_{a_1' a_{\sigma (1)}}^{\gamma_1' \gamma_{\sigma (1)}} (t_1 ', t_{\sigma(1)} ) \dots \Delta_{a_k' a_{\sigma (k)}}^{\gamma_k' \gamma_{\sigma (k)}} (t_k ', t_{\sigma(k)} )
}
\eeq
\end{widetext}

This series can be sampled using stochastic sampling techniques \cite{muhlbacherRabaniPRL2008,schiroFabrizioPRB2009,wernerMillisPRL2006} or approximately resummed  \cite{
aokiWernerRMP2014,rueggMillisPRB2013}. For both purposes, it is useful to define the Feynman rules for the series \eqref{eq:hybExp}.

\subsubsection{Matrix representation}
Each term of the hybridization expansion \eqref{eq:hybExp} must be understood as a composition of applications of super-operators, from the right-most to the left-most one, on the initial density operator.
We remark that in the usual representation in which states of the Hilbert space are vectors and operators are matrices, super-operators are rank-4 tensors. Instead, if we write operators as vectors, then super-operators become matrices. 
This is convenient to evaluate terms of the hybridization expansion \eqref{eq:hybExp} as matrix products.
Namely 
\beq
\es{
&\mcV_0(t,t_k') \mcD^\da_{a_k '(t_k ' \gamma_k' )} \dots \mcD_{a_1 (t_1 \gamma_1)} \mcV_0 (t_1,0) \rho_{I}(0) \\ \rw \, &  \dbar{\mcV}_0(t,t_k') \dbar{\mcD}^\da_{a_k '(t_k ' \gamma_k' )}  \dots  \dbar{\mcD}_{a_1 (t_1 \gamma_1)} \dbar{\mcV}_0 (t_1,0) \ket{\rho_{I}(0)}
}
\eeq
using double bars to indicate matrices.
Operators are represented as vectors in a space which is the tensor product (indicated with $\otimes$) of two copies of the original Hilbert space. The vectorization procedure, taking the example of the density operator, reads:
\beq 
\label{eq:opToVec} 
\rho = \sum_{n,m} \rho_{n m } \ket{n} \bra{m} \rw \sum_{n,m} \rho_{n m } \ket{n} \otimes \ket{m} \equiv \ket{\rho }
\eeq
The matrices $\dbar{\mathcal{V}},\dbar{d},\dbar{d}^\da$ corresponding to the super-operators $\mcV,d,d^\da$ are defined by the following simple procedure. Let's consider the super-operator $\mathcal{S} \bullet  = A \bullet B $. Representing $\rho$ as a vector $\ket{ \rho}$, then also $\mathcal{S} \rho $ will be a represented as a vector according to eq. \eqref{eq:opToVec}. It's simple to show that $\ket{\mathcal{S} \rho} = \dbar{\mathcal{S}} \ket{\rho } $ defining $\dbar{\mathcal{S}}=\dbar{A} \otimes \dbar{B}^T$.
The matrix form of $\mcD_\pm,\mcD_\pm^\da$ in the doubled Hilbert space is
\begin{align*}
 \dbar{\mcD}_+ &= \dbar{d} \otimes \dbar{\id} & \dbar{\mcD}_+^\da &= \dbar{d}^\da \otimes \dbar{\id} \\ 
 \dbar{\mcD}_- &= \dbar{\id} \otimes \dbar{d}^T &  \dbar{\mcD}_-^\da &= \dbar{\id} \otimes \dbar{d}^* \\ 
\end{align*}
with $\dbar{\mcD}_\pm^\da$ the hermitian conjugate of $\dbar{\mcD}_\pm$. 
These rules allow to obtain the matrix representation of the Liouvillian and then of $\mcV_0$.

\subsection{Feynman Rules}

\begin{figure}
\includegraphics[scale=0.35]{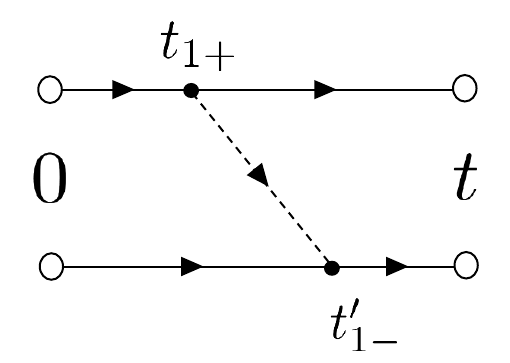} 
\caption{The Feynman diagram representing eq. \eqref{eq:sampleDiagram} \label{fig:sampleDiagram} }
\end{figure}

\ig[fig:feynmanRules]{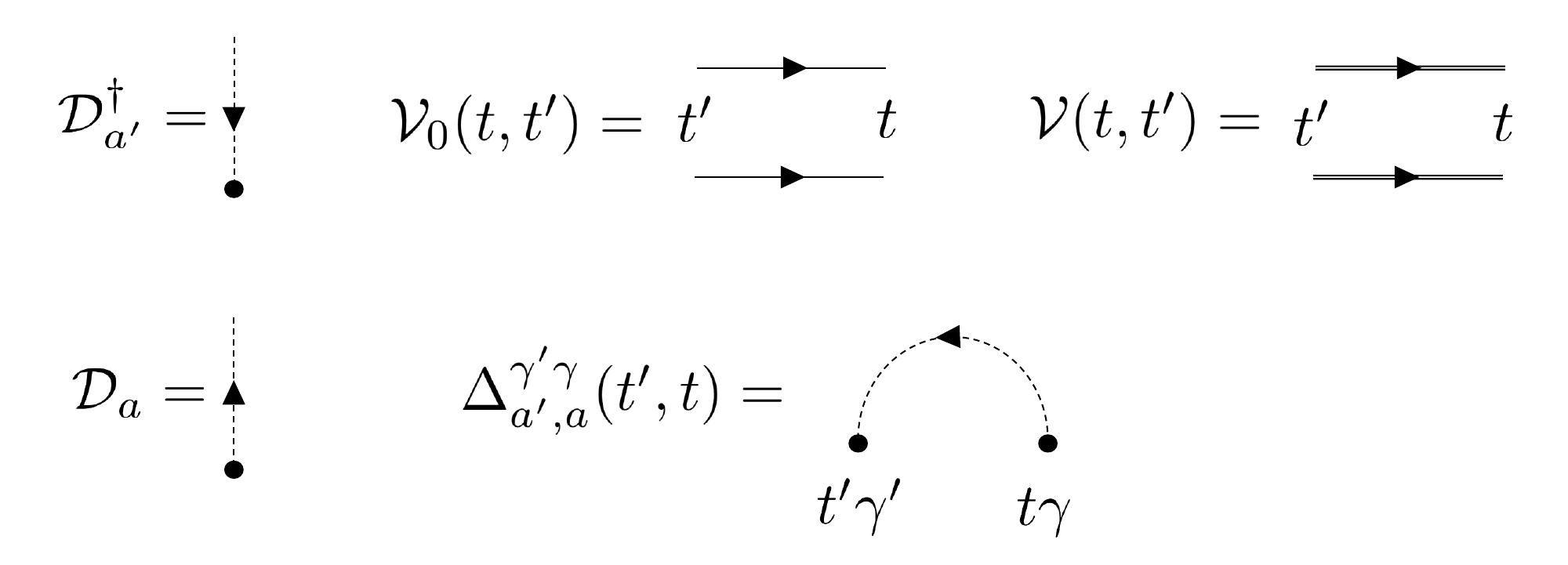}{1}{The Feynman rules to represent the hybridization expansion \eqref{eq:hybExp}.
The arrow of the hybridization line $\Delta$ goes from a $\mcD$ super-operator to the first a $\mcD^\da$ one.} 

the Feynman rules to draw the hybridization expansion \eqref{eq:hybExp} are represented in figure \ref{fig:feynmanRules}.
We will use these rules to draw a term with $2k$ annihilation and creation super-operators, with a particular ordering for the times $\{ t_i, t_i' \dots t_1,t_1' \} $ and a choice of a permutation $\{ \sigma (1), \sigma (2) \dots \sigma (k) \}$.
To do that, we draw a couple of parallel axes representing the double contour from time $t=0$ to time $t$.
$\mcD_\gamma$ ($\mcD^\da_\gamma$) super-operators are represented as a dashed half-line with outwards (inwards) arrows, stemming from the contour branch $\gamma$.
The dashed half-lines corresponding to the super-operators $\mcD_{a_i ' ( t_i' \gamma_i')}^\da $ and $\mcD_{a_{\sigma (i)}  ( t_{\sigma (i)} \gamma_{\sigma (i)})} $ are joined together to form a hybridization line, representing the hybridization function $\Delta^{\gamma_i' \gamma_{\sigma (i)}}_{a'_i a_\sigma (i)}(t_i',  t_\sigma (i)) $, which has an arrow going from $\mcD$ to $\mcD^\da$.
Then, each part of the double contour between two integration times, drawn as two parallel solid segments, represents a time-propagation super-operator $\mcV_0$. The dressed evolution operator $\mcV$ is drawn by replacing the contour solid lines by double lines.
%
%
As an example, the diagram corresponding to \beq
\label{eq:sampleDiagram}
 i \int_0^t  dt_1 \int_0^{t_1}  dt_1' \mcV_0(t,t_1') \mcD ^\da_{a_1' -} \mcV_0(t_1',t_1) \mcD_{a_1 +} \mcV_0(t_1,0) \Delta_{a_1', a_1}^{- +} (t_1' ,t_1 )  \eeq is shown in figure \ref{fig:sampleDiagram}.
 All the diagrams with $2k$ annihilation and creation super-operators are generated by connecting $\mcD^\da$ super-operators to $\mcD$ ones in all possible choices of permutations $\sigma$ and considering all possible time orderings of integration times. 
\section{Self-Consistent Diagrammatic Resummations Techniques}

In this section we start from the hybridization expansion derived in section III, which involves bare diagrams to all orders, and use the Feynman rules to introduce diagrammatic resummation techniques.

To proceed further it is useful to draw more compact diagrams where the double contour is collapsed on a single time-axis and thus a time propagation $\mcV_0$ is represented by a single line, as we show in figure \ref{fig:collapsedContour}. 
These compact diagrams represent an ensemble of diagrams drawn with the rules we introduced in \ref{fig:feynmanRules}.
\ig[fig:collapsedContour]{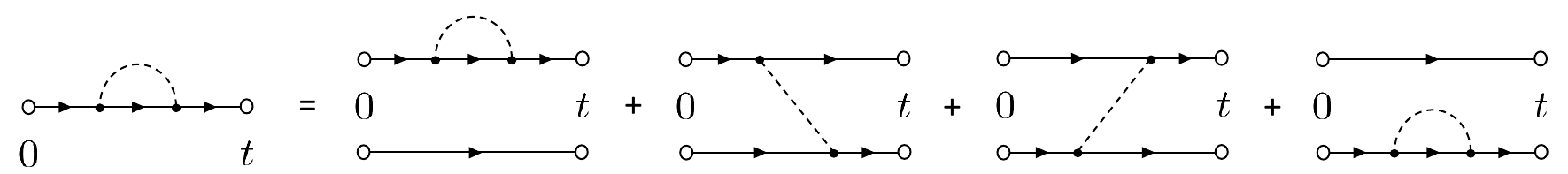}{1}{Compact diagrams represent an ensemble of diagrams hiding the double contour structure. By omitting the arrows on hybridization lines, we mean that all the possible choices must be considered.}
The advantage of this notation is that all the diagrams represented by a single compact diagram have the same topology in terms of being 1-particle irreducible or non-crossing. Figure \ref{fig:1PI_diagrams} shows the hybridization expansion drawn using these compact diagrams.

\subsection{Dyson Equation}
%


As a first step it is useful to distinguish diagrams which are one-particle irreducibles, i.e.  compact diagrams which cannot be separated, by cutting a solid line, in two parts that are not connected by any hybridization line, as indicated in figure \ref{fig:1PI_diagrams}. Then, we introduce the self-energy $\Sigma$ as the sum of one particle irreducible (1PI) diagrams.
All the non-1PI diagrams can be obtained by joining some 1PI diagrams with solid lines, thus the whole series can be written as $$\mcV  = \mcV_0 + \mcV_0 \circ \Sigma \circ \mcV_0 + \mcV_0 \circ \Sigma \circ \mcV_0 \circ \Sigma \circ \mcV_0  + \dots$$



\ig[fig:1PI_diagrams]{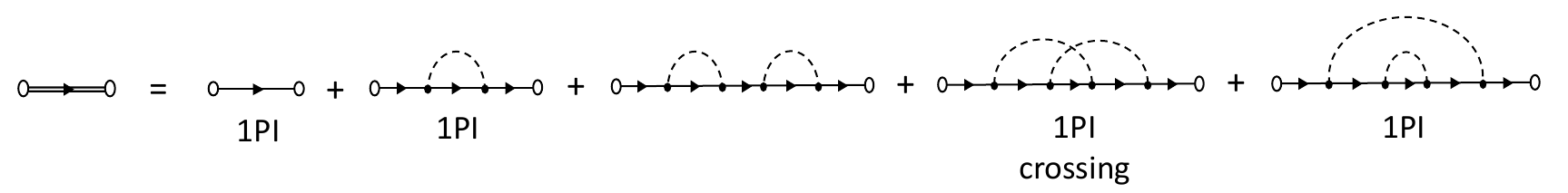}{1}{1PI diagrams of the hybridization expansion in eq. \eqref{eq:hybExp}}

We remark that the objects composing this series, $\mcV_0$ and $\Sigma$, are super-operators and the series must be understood as a composition of applications of super-operators, from the right-most to the left-most one, on a target operator. Self-energies and propagators are joined by the circle operation\cmm{, $\circ$,} standing for a super-operator application and a partial time convolution. Using brackets to stress that we refer to a super-operator application and the symbol $\bullet$ to indicate a target operator, we have $$ \Sigma(t,t_1)  \circ \mcV_0(t_1,t') \equiv \int_{t'}^t dt_1 \Sigma(t,t_1) \lsq \mcV_0(t_1,t')  \lsq \bullet \rsq\rsq$$
\cmm{The series above sums up to the Dyson equation $\mcV  = \mcV_0 + \mcV_0 \circ \Sigma \circ \mcV =  \mcV_0 + \mcV \circ \Sigma \circ \mcV_0 $, or equivalently, in integro-differential form}

\beq
\label{eq:dysonNCA} 
\cmm{
\pt \mcV (t,t') = \mathcal{L} \mcV(t,t') + \int_{t'}^t dt_1   \Sigma(t,t_1) \mcV (t_1,t') }
\eeq 
\cmm{When the self-energy of the non-Markovian environment $\Sigma$ is set to zero, this equation yields $\mathcal{V}(t) = e^{\lind t}$, which is the Lindblad evolution.}

\cmm{
One of the main effects of dissipative dynamics is that the system may forget about initial conditions and reach the same stationary state for any initial condition. 
Assuming a stationary state exists for a non-Markovian map $\mathcal{V}$ defined by the Dyson equation \eqref{eq:dysonNCA}, then it satisfies 
\beq
\label{eq:ss_equation}
\lp \mathcal{L} +  \int_{0}^\infty dt_1 \Sigma(\infty,t_1) \rp \rho_{ss} = 0 
\eeq 
Setting the non-Markovian self-energy to zero, this equation reduces to the Lindblad condition for the stationary state.
The derivation of this equation is in the supplementary material.
}
\subsection{The Non-Crossing Approximation}
\label{sec:eqNCA}

The non-crossing approximation (NCA) corresponds to approximating the series for $\mcV $, and thus also for $\Sigma$, by considering only the compact diagrams in which the hybridization lines do not cross \cite{Bickers1987,Nordlander1999,Eckstein_NCA_PRB10,rueggMillisPRB2013}. 
The NCA diagrams composing the self-energy are shown in figure \ref{fig:resummedSigmaExt}.
\ig[fig:resummedSigmaExt]{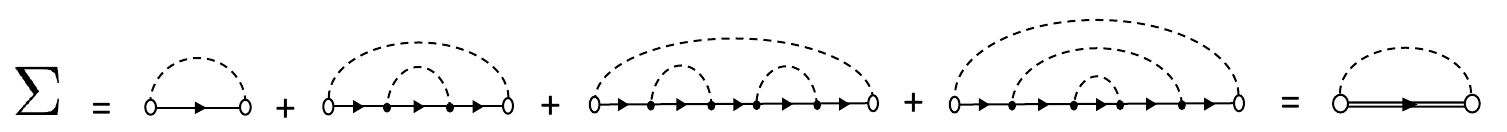}{1}{The NCA series of the self-energy $\Sigma$. The resummed series for $\Sigma$ corresponds to its $k=1$ diagrams, where the bare propagator $\mcV _0$ is replaced with the dressed one $\mcV $.}
In order to prove the second equality in figure \ref{fig:resummedSigmaExt}, we remark that the first and last times of a self-energy diagram must be connected together by an hybridization line. If it's not the case, in fact, the resulting diagram is either non-1PI or it's crossed.
Then all the diagrams of $\Sigma$ (in NCA) are obtained connecting the intermediate times to form all the possible non-crossing diagrams, (not only the 1PI ones this time). But the latter diagrams in turn define the NCA series for $\mcV $. This proves the second equality in figure \ref{fig:resummedSigmaExt}. We remark that then, the NCA self-energy coincides with its contributions (with $k=1$), where the bare propagator $\mcV_0$ is replaced with the dressed one $\mcV $.

To obtain an analytic expression of the self-energy, we have to cast the $k=1$ term of the hybridization expansion \eqref{eq:hybExp} in a form in which the innermost integration time is lower than the outermost, that is with integrals of the form  $ \int_{t'}^t dt_1 \int_{t'}^{t_1} dt_2$. In doing so, one must deal with the signs coming out of the time ordering. We report the calculation in the supplementary material.


The expression for the self-energy eventually reads
\begin{widetext}
\begin{align}
\label{eq:selfEnergyExplicit}
&\Sigma (t_1,t_2) =  \sum_{a , b} \sum_{\alpha \beta \in \{ + , - \}  }   - \alpha^{(1+\xi)/2} \beta i \lsq  \Delta_{b  a }^{\beta \alpha} (t_1, t_2 ) \mcD_{ \beta  b }^\da \mcV (t_1,t_2) \mcD_{ \alpha a } + \xi  \Delta_{ a b }^{\alpha \beta} (t_2,t_1 ) \mcD_{\beta b} \mcV (t_1,t_2) \mcD^\da_{\alpha a}   \rsq 
\end{align}
\end{widetext}
where $\alpha, \beta \in \{ + , - \}$ are contour indices, $a,b$ are the fermionic generic indices. 

We can interpret the two terms in eq. \eqref{eq:selfEnergyExplicit} as follows.
The first term propagates a hole in the impurity (applies $\mcD^\da$ first and then $\mcD$) and a particle in the bath, the latter being described by a hybridization function with the same time arguments of $\mcV$;
The second term propagates a particle in the impurity and a hole in the bath with a hybridization function with opposite time arguments than $\mcV$.
%
%

\cmm{Few comments are in order here, concerning the above result. First, in absence of the Markovian environment, that is by replacing $\mcV (t,t')$ and $\mcV _0(t,t')$ with $\mcU(t,t')$ and $\mcU_0(t,t')$, our results are equivalent to non-equilibium NCA schemes for unitary dynamics \cite{Eckstein_NCA_PRB10,strandWernerPRX2015,peronaciPRL2018}. 
 There is a formal difference consisting in our super-operators formulation of the hybridization expansion and of the Dyson equation, that is necessary to consider the additional Markovian environment without further approximations, but this difference is only formal and does not affect the results.
In fact\cmm{, if $N$ is the dimensionality of the impurity Hilbert space,} the usual non-equilibrium NCA propagator \cite{strandWernerPRX2015,peronaciPRL2018} has different Keldysh components, each of them being an $N \times N$ matrix, while our $\mcV$ propagator is a $N^2 \times N^2$ matrix with no Keldysh components. }

\cmm{Furthermore, the result obtained for the NCA self-energy in Eq.~(\ref{eq:selfEnergyExplicit}) makes clear that for a bath hybridization which is delta-correlated in time the resulting self-energy contribution to the Dyson equation takes the form of an additional Lindblad dissipator. In other words, one can recover the Lindblad master equation from our diagrammatic NCA approach in the Markovian limit, and possibly discuss corrections to the master equation from higher order terms, as recently done \cite{Muller2017}. }


\subsection{Properties of the NCA propagator}
\label{subsec:propNCA}
\begin{figure}
\epsfig{figure=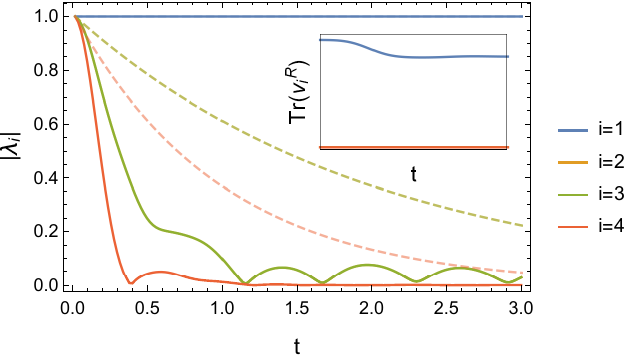,scale=0.8}
\caption{Real-time evolution of the absolute value of the eigenvalues of the impurity propagator. While an eigenvector with eigenvalue one is present at all times, all the other eigenvalues decay to zero at long times. The $i=2$ and $i=3$ curves coincide because the corresponding eigenvalues are complex conjugates. The decay is purely exponential for a Markovian system while strong deviations appear in the non-Markovian case. \cmm{The inset shows that the right eigenstates of $\mcV (t)$ with different-from-one eigenvalues are traceless, while the right eigenstate with eigenvalue one has a finite (unnormalized) trace.} Parameters: $\epsilon_0 = 5$, $\gamma = \gamma_l = \gamma_p = \gamma_d = 0.5$, $w = 10$, $\eta=1$, $\Delta t = 0.02$, $\rho_0 =\ket{0} \bra{0}$. }
\label{fig:eigOfV}
\end{figure}

The propagator $\mcV (t,t')$ obtained in NCA is the time-evolution super-operator of the reduced density operator of the impurity. Assuming to switch on the interaction with the \cmm{baths} at time $t=0$, then the density operator of the impurity at time $t$ is given by $\rho_I(t) = \mcV(t,0)  \rho_I(0) $.
A time evolution super-operator must be a convex-linear, completely positive and trace-preserving map \cite{breuerPetruccione2010}. It is natural to ask which of these properties are preserved by the NCA approximation. $\mcV $ is a obviously a linear map, implying it is also convex linear. We proved that it is trace-preserving and that it preserves hermiticity \cmm{(see supplementary material), while proving or disproving whether the map is completely positive is a tough task \cite{Reimer2018} that will be addressed in the future.}
We stress that $\mcV (t,t')$ describes a non-Markovian evolution, so it does not form a semi-group, that is $\mcV (t,t') \neq \mcV(t,t_1) \mcV(t_1,t')$ with $t'<t_1<t$. 
Time-evolution super-operators have also interesting spectral properties following from trace preservation. We refer to the supplementary material for the proof. We call $\lambda_i(t,t'), v^R_i (t,t')$ the eigenvalues and right eigenvectors of $\mcV (t,t')$, depending on time.
As it preserves the trace, $\mcV (t,t')$ must have at least one eigenvalue equal to one, say $\lambda_0 \equiv 1$. 
\cmm{If we assume this eigenvalue is non-degenerate, then all the others eigenvectors with $i \neq 0 $, are traceless. }
\cmm{As a consequence of these properties, if one evolves an initial state $\rho(0)$ and expands $\rho(t)$ on the instantaneous eigenvectors $v^R_i (t,0)$, then those eigenvectors with $i \neq 0$ will represent decay modes of the dynamics as they will be suppressed by their corresponding vanishing eigenvalues for long times, while $v^R_{i=0} (t,0)$ will evolve in time undumped until reaching a stationary value, representing the stationary state of the non-Markovian evolution.}
We will \cmm{numerically} check these properties in figure \ref{fig:eigOfV} of the next section, where we will apply our NCA algorithm to a specific example. 

\begin{figure*}
\centering
\begin{minipage}[t]{.4\textwidth}
\epsfig{figure=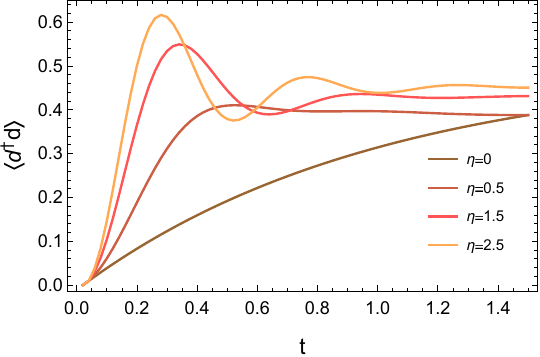,scale=0.7}
\epsfig{figure=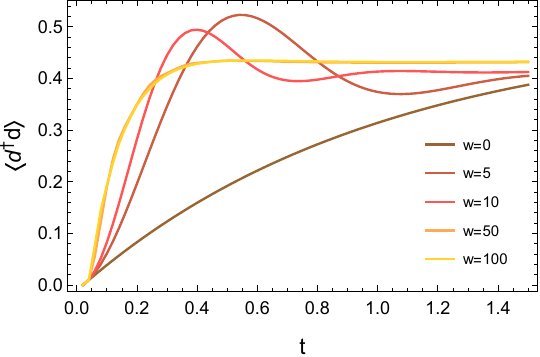,scale=0.7}
\end{minipage}\qquad
\begin{minipage}[t]{.4\textwidth}
\epsfig{figure=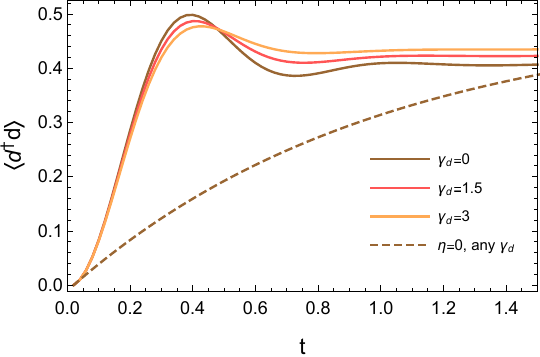,scale=0.7}
\epsfig{figure=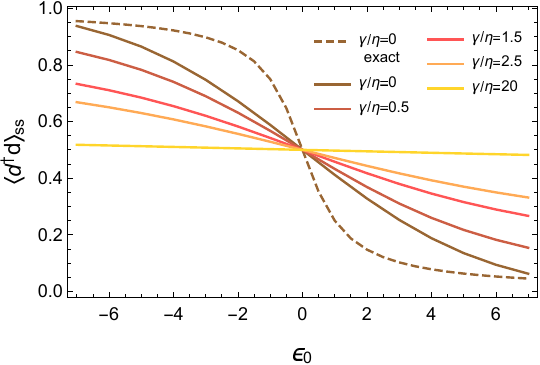,scale=0.7}
\end{minipage}
\caption{Dynamics of the number of fermions for different sets of parameters, namely changing the hybridization strength (Top left) the bandwidth (bottom left) and the dephasing rate (top right). Average population of the stationary state as a function of the energy level (bottom right). Parameters: $\epsilon_0 = 1$, $\gamma = \gamma_l = \gamma_p = \gamma_d = 0.5$, $w = 10$, $\eta=1$, $\Delta t = 0.02$, $\rho_0 =\ket{0} \bra{0}$. }
\label{fig:results4Panels}
\end{figure*}

\section{Case study: spin-less fermionic impurity}

As a non-trivial application of the NCA approach for open system described so far, we consider here a model of a single-mode, spin-less fermionic impurity with Markovian losses, pump and dephasing and further coupled to a non-Markovian fermionic environment, as described in equation \eqref{eq:bathCoupling}. \cmm{We notice that the model in absence of dephasing, also known as Resonant Level Model, is quadratic in all the fermionic degrees of freedom and therefore easily solvable, with analytical expressions known for the wide band limit. At finite dephasing this is no longer the case and the model cannot, to the best of our knowledge, be solved by simple means. This could be understood naturally in the Keldysh approach, where the dephasing would result in density-density type of coupling between different Keldysh branches. }

The Markovian dynamics is described by a Lindblad master equation
\begin{align*}
\pt \rho_I^0 &= \mathcal{L} \rho_I^0 \\ 
\mathcal{L} \rho_I^0 &= - i \lsq H_I, \rho_I^0 \rsq + \lp \gamma_l \mathcal{D}_l + \gamma_p \mathcal{D}_p + \gamma_d \mathcal{D}_d \rp \rho_I^0 \\
H_I &= \epsilon_0 d^\da d \\
 \mathcal{D}_l \rho_I^0 &= d \rho_I^0 d^\da - \oh \{ d^\da d, \rho_I^0 \} \\ 
 \mathcal{D}_p \rho_I^0 &= d^\da \rho_I^0 d - \oh \{ d d^\da, \rho_I^0 \} \\ 
  \mathcal{D}_d \rho_I^0 &= d^\da d \rho_I^0 d^\da  d - \oh \{ d^\da d, \rho_I^0 \}
\end{align*}
where $ \epsilon_0$ is the energy of the fermionic level. 

The effect of the non-Markovian environment on the impurity is completely determined by its hybridization function \eqref{eq:hybridizationFunction}. Here we choose a zero temperature, particle hole symmetric, fermionic bath with constant density of states of bandwidth \cmm{$2w$} and with coupling strength to the impurity $\eta$. In this case the hybridization function depends only on time-differences.
%
As a consequence, one can show that also $\mcV$ and $\Sigma$ depend only on time differences and we will set $t'=0$. With these definitions we get for the hybridization functions
\begin{align*}
\Delta^{+- }(t) &=  2 i \eta\, e^{i w t / 2 } \sin \lp  w \, t / 2 \rp / t\\ 
\Delta^{-+ }(t) &=  -2 i \eta\, e^{-i w t / 2 } \sin \lp  w \,  t / 2 \rp / t \\ 
\end{align*}
To solve the Dyson equation \eqref{eq:dysonNCA} numerically
we use the simple discretization scheme $\partial f(t) = \lsq f(t+\Delta t ) - f(t) \rsq / \Delta t$, $\int_0^t dt_1 f(t_1) = \Delta t/2 \sum_{l=0}^{t/\Delta t -1 } \lsq f( (l+1) \Delta t ) +  f( l \Delta t )  \rsq $, with time-step $\Delta t$. More refined integration methods are explained in detail in \cite{aokiWernerRMP2014}.
The Hilbert space of the impurity has size $N = 2$, so that the super-operator $\mcV $ has size $N^2 \times N^2 = 4 \times 4$ .

\subsection{Results}

We start analyzing the spectral properties of the propagator $\mcV (t)$, which have been discussed generically in the previous section, and are reported in figure \ref{fig:eigOfV}. In \cmm{the main panel} we plot the time dependence of the \cmm{absolute value of the} eigenvalues of $\mcV (t)$, both in the purely Markovian case (dashed lines) as well in presence of both kind of dissipations. In both cases there is an eigenvalue which remains equal to one, while the others decay to zero at long times, as pointed out in section \ref{subsec:propNCA}. However the nature of this decay is rather different in the two cases, showing a faster dynamics and long time oscillations in the non-Markovian case as opposed to a pure exponential decay in the Markovian one. \cmm{The inset of figure \ref{fig:eigOfV} shows instead that all the right eigenstates of $\mcV (t)$ with different-from-one eigenvalues are traceless, while the right eigenstate with eigenvalue one has a finite trace (that we could normalize to one at every time).}

\begin{figure}
\epsfig{figure=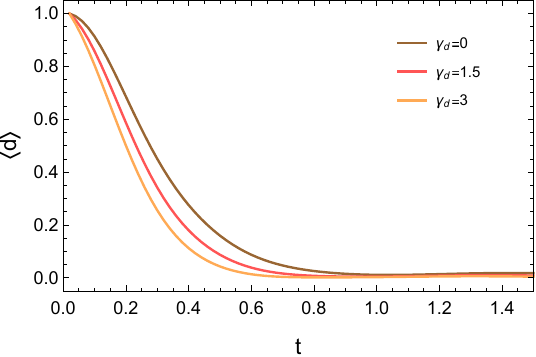,scale=0.7}
\caption{Dynamics of the density matrix coherence for different values of the dephasing. Parameters: $\epsilon_0 = 1$, $\gamma = \gamma_l = \gamma_p = \gamma_d = 0.5$, $w = 10$, $\eta=1$, $\Delta t = 0.02$, $\rho_0 =\ket{0} \bra{0} + \ket{0} \bra{1} + \ket{1} \bra{0}$. }
\label{fig:coherence}
\end{figure}

We then consider the dynamics of a simple observable, such as the density of fermions in the impurity level, as a function of time and for different parameters (see panel figure \ref{fig:results4Panels}). In the left figures we plot the dynamics for different values of the coupling $\eta$ (top) and bandwidth $w$ (bottom) of the non-Markovian environment, in presence of fixed Markovian losses, pump and dephasing. We see that with respect to the purely Markovian dynamics, characterized by a simple exponential relaxation, the NCA approach captures aspects related to the non-Markovian nature of the environment. In particular the dynamics becomes characterized by oscillations whose amplitude and frequency increase with the coupling $\eta$. \cmm{Similarly, increasing the bandwidth of the non-Markovian environment reduces the oscillations in the population dynamics, which disappear in the large bandwidth limit, as it is the case for the unitary dynamics of the Resonant Level Model. This is not surprising, since oscillations at short-times $t\sim 1/w$ come from high-energy modes of the bath.}

Overall the Non-Markovian environment makes the dynamics substantially faster. 
\cmm{In the top-right plot we discuss the role of dephasing, that is actually very interesting as it shows an effect of the combined Markovian and non-Markovian environments. The dashed line shows that, for Markovian dissipation only, the dephasing does not affect the population dynamics; this is well understood as the dephasing dissipator commutes with the number operator. It's interesting to see that, instead, combined with a non-Markovian environment the dephasing has an impact on the dynamics of populations; this is a smaller effect as it involves both the Markovian and the non-Markovian environments.
This effect can be understood as follows: let's consider a process in the time-evolution in which the non-Markovian environment applies a $d^\da_+$ operator on the density matrix and then a $d_+$ operator after some time. The application of the creation operator converts the populations of the density matrix into coherences. Then the Markovian dephasing dumps those coherences, which are then converted back to populations when the non-Markovian environment applies the annihilation operator. As a net effect, the dephasing has produced a change into populations, as it is shown in the top-right plot. We also note that not only the dynamics, but also the stationary values of the occupation change with the dephasing. A more direct effect of the dephasing appears in the coherences, i.e. in the off-diagonal elements of the density matrix, which decay to zero faster as $\gamma_d$ is increased, as we show in figure~\ref{fig:coherence}. 
 }
For what concerns the stationary state, we notice from the bottom-right plot that the average density would be independent of the  \cmm{energy of the fermionic level in the purely Markovian case, which leads to a infinite-temperature fully-mixed stationary state for the chosen dissipation rates.
This makes sense as a Markovian bath has no energy structure, thus the level effectively sees always the same bath even if it's shifted in energy.}
On the other hand the coupling to the non-Markovian bath makes the population depend strongly on the position of the energy level and gives a result which is in good agreement with exact analytical calculations (dashed line); to justify the quantitative discrepancy with this analytical result, we stress that for the non-interacting model we consider here the NCA approximation, which is based on a strong coupling expansion, is not expected to be exact. 
%
%


\section{Conclusions}

In this work we have focused on a model for a quantum impurity coupled simultaneously to a Markovian and a non-Markovian environment. We derived a formal hybridization expansion for the evolution super-operator of the impurity, obtained after tracing out all the bath degrees of freedom. This result generalizes to non-unitary, Markovian case the hybridization expansion obtained for unitary quantum impurity models. As such it provides the natural starting point for the development of stochastic sampling techniques of the dissipative real-time dynamics of the impurity based on Diagrammatic Monte Carlo, that we leave for future studies. 

Starting from this expansion we define real-time diagrammatic rules and write down a Dyson Equation for the impurity propagator that we evaluate retaining only non-crossing diagrams, an approximation which is known to capture some aspects of the impurity physics at strong coupling. The resulting approach leads to a trace and hermiticity preserving Non-Markovian dynamical map, with consequences on the spectral properties of the evolution super-operator, while proving its complete positivity in full generality remains an open question. 

\cmm{It is interesting to comment on the relation between our approach and related methods to deal with impurity models coupled to multiple baths. While in principle both the hybridization expansion~\cite{GolezEtAlPRB15} as well as the strong coupling diagrammatic resummation~\cite{ThetaEtAlPRB16} can be generalized in presence of multiple environments, taking the Markovian limit from the start has some practical and conceptual advantage. In particular, we can take direct advantage of the local nature of the Markovian evolution and perform an expansion around an \emph{atomic-limit} which now contains not only interaction but also drive and dissipation. This limit can be solved exactly by direct diagonalization of a Lindbladian, as opposed to treating exactly the degrees of freedom of the environment which are not expected to introduce new physics in presence of memory-less Markovian correlations. The key idea is therefore to treat on equal footing all the energy scales related to fast processes, while resorting to perturbation theory when dealing with processes leading to slowing decay correlations such as the coupling to gapless reservoirs.}

As an application, we solved numerically the Dyson equation for the simple model of a fermionic, single-mode impurity, with Markovian losses, pump and dephasing and a non-Markovian, zero temperature environment. This model is non-trivial for the presence of dephasing, which is a quartic term in fermionic operators. This simple implementation allowed to check the spectral properties of the evolution super-operator and to study how Markovian dynamics gets modified by coupling to a non-Markovian environment. 
\cmm{In particular our method allowed to show a physical consequence of coupling simultaneously to Markovian and non-Markovian environments: Markovian dephasing combined with non-Markovian processes leads to a change in impurity occupations.}
Future directions include the exploration of more complex impurity models involving internal degrees of freedom such as the Anderson Impurity model as well as bosonic extensions and to use of this method as an impurity solver within a dynamical mean field theory approach to driven-dissipative systems.

\section{Supplementary Material}

The Supplementary Material includes (i) a derivation of the equation for the steady state density matrix starting from the Dyson Equation for the non-Markovian map (ii) the detailed derivation of the NCA self-energy and (iii) the proofs of trace and hermiticity preservation of the NCA dynamics.

\pagebreak

\onecolumngrid
\begin{center}
  \textbf{\large Quantum Impurity Models coupled to Markovian and Non Markovian Baths\\Supplementary Material}\\[.2cm]
(Dated: \today)\\[1cm]
\end{center}

\setcounter{equation}{0}
\setcounter{figure}{0}
\setcounter{table}{0}
\setcounter{page}{1}
\renewcommand{\theequation}{S\arabic{equation}}
\renewcommand{\thefigure}{S\arabic{figure}}
\renewcommand{\bibnumfmt}[1]{[S#1]}
\renewcommand{\citenumfont}[1]{S#1}

\section{Stationary state}
{
Assuming a stationary state exists for a non-Markovian map $\mathcal{V}$ defined by the Dyson equation 
\beq
\label{eq:dyson}
\pt \mcV (t,t') = \mathcal{L} \mcV(t,t') + \int_{t'}^t dt_1   \Sigma(t,t_1) \mcV (t_1,t') 
\eeq 
then, setting $t'=0$, it satisfies 
\beq
\label{eq:ss_equation}
\lp \mathcal{L} +  \int_{0}^\infty dt_1 \Sigma(\infty,t_1) \rp \rho_{ss} = 0 
\eeq 
In order to derive this equation, we focus on the propagator $V(\infty,0)$, that projects any initial state on the stationary state: $\rho_{ss} =V(\infty,0) \rho(0) $.
This propagator obeys the Dyson equation 
$$
\lim_{t\rw \infty }\pt \mcV (t,0) = \mathcal{L}\mcV (\infty,0) + \int_{0}^{\infty} dt_1   \Sigma(\infty,t_1) \mcV (t_1,0)
$$
At this point we need to make some assumptions based on physical arguments. $\mathcal{V}(t_1,0)$ is expected to have a transient dynamics in a finite time interval of duration $t_\text{tr}$, starting at time $t_1=0$, and then to become stationary, i.e.  $\lim_{t\rw \infty }\pt \mcV (t,0) =0$.
In addition, the system is supposed to lose memory of initials conditions, thus the convolution in the above Dyson equation must be cutoffed by the self-energy $\Sigma(\infty,t_1)$ for $\infty-t_1> t_\text{mem}$. 
Then, in the region where $\Sigma(\infty,t_1)$ is non-zero, $ \mathcal{V}(\infty,t_1)$ is stationary and we can replace it with $ \mathcal{V}(\infty,0)$. With these arguments we get $\lp \mathcal{L} +  \int_{0}^\infty dt_1 \Sigma(\infty,t_1) \rp  \mathcal{V}(\infty,0 ) = 0 $ and applying it to any initial state we find \eqref{eq:ss_equation}.
}

\section{Derivation of NCA self-energy}
To obtain an analytic expression of the self-energy, we have to cast the $k=1$ term of the hybridization expansion obtained in the main text in a form in which the innermost integration time is lower than the outermost, that is with integrals of the form  $ \int_{t'}^t dt_1 \int_{t'}^{t_1} dt_2$. In doing so, sign of each contribution will be determined by the time-ordering $T_C$. 
The $k=1$ term of the hybridization expansion, omitting the fermionic indices to simplify the expressions, reads
\beq
\label{eq:k1Term}
\es{
i \mcV^{(1)} (t,0)= & \sum_{\gamma_1,\gamma_2} \gamma_1 \gamma_2  \int_0^t  dt_1 \int_0^t dt_2  T_F T_C  \mcV_0(t,t_1) \mcD^\da_{(t_1 \gamma_1 )} \mcV_0(t_1, t_2) \mcD_{ (t_2 \gamma_2)} \mcV_0 (t_2,0)  \Delta^{\gamma_1 \gamma_2} (t_1, t_{2} ) = \\ 
= &   \int_0^t  dt_1 \int_0^t dt_2  T_F T_C  \mcV_0(t,t_1)  \mcD^\da_{(t_1 + )} \mcV_0(t_1, t_2)  \mcD_{ (t_2 +)} \mcV_0 (t_2,0)  \Delta^{++} (t_1, t_{2} ) + \\ 
 &-  \int_0^t  dt_1 \int_0^t dt_2  T_F T_C  \mcV_0(t,t_1)  \mcD^\da_{(t_1+ )} \mcV_0(t_1, t_2)  \mcD_{ (t_2 -)} \mcV_0 (t_2,0)  \Delta^{+-} (t_1, t_{2} ) + \\ 
 &-    \int_0^t  dt_1 \int_0^t dt_2  T_F T_C  \mcV_0(t,t_1)  \mcD^\da_{(t_1- )} \mcV_0(t_1, t_2)  \mcD_{ (t_2+)} \mcV_0 (t_2,0)  \Delta^{-+} (t_1, t_{2} ) + \\ 
&  \int_0^t  dt_1 \int_0^t dt_2  T_F T_C  \mcV_0(t,t_1)  \mcD^\da_{(t_1- )} \mcV_0(t_1, t_2)  \mcD_{ (t_2 -)} \mcV_0 (t_2,0)  \Delta^{--} (t_1, t_{2} ) 
}
\eeq
where we have summed on contour indices $\gamma_1$ and $\gamma_2$.
Now we brake the integrals in two pieces, for $t_2>t_1$ and $t_2<t_1$, which allows to put the operators in a time-ordered fashion according to $T_C$. 
\beq
\es{
& i \mcV^{(1)} (t,0) =\\ =&   \int_0^t  dt_1 \int_0^{t_1} dt_2  T_F   \mcV_0  \mcD^\da_{(t_1 + )} \mcV_0  \mcD_{ (t_2 +)} \mcV_0   \Delta^{++} (t_1, t_{2} ) +\xi   \int_0^t  dt_1 \int_{t_1}^{t} dt_2  T_F   \mcV_0  \mcD_{ (t_2 +)} \mcV_0  d^\da_{(t_1 + )} \mcV_0   \Delta^{++} (t_1, t_{2} ) + \\ 
 &-\xi  \int_0^t  dt_1 \int_0^{t_1} dt_2  T_F   \mcV_0   \mcD_{ (t_2 -)} \mcV_0  \mcD^\da_{(t_1+ )} \mcV_0   \Delta^{+-} (t_1, t_{2} ) -\xi   \int_0^t  dt_1 \int_{t_1}^{t} dt_2  T_F   \mcV_0   \mcD_{ (t_2 -)}  \mcV_0  \mcD^\da_{(t_1 + )} \mcV_0   \Delta^{+-} (t_1, t_{2} ) + \\ 
 &-  \int_0^t  dt_1 \int_0^{t_1} dt_2  T_F   \mcV_0  \mcD^\da_{(t_1- )} \mcV_0  \mcD_{ (t_2+)} \mcV_0   \Delta^{-+} (t_1, t_{2} )  -  \int_0^t  dt_1 \int_{t_1}^{t} dt_2  T_F   \mcV_0  \mcD^\da_{(t_1 - )} \mcV_0  \mcD_{ (t_2 +)} \mcV_0   \Delta^{-+} (t_1, t_{2} ) + \\ 
&+\xi  \int_0^t  dt_1 \int_0^{t_1} dt_2  T_F   \mcV_0  \mcD_{(t_2- )} \mcV_0  \mcD^\da_{ (t_1 -)} \mcV_0   \Delta^{--} (t_1, t_{2} )  +   \int_0^t  dt_1 \int_{t_1}^{t} dt_2  T_F   \mcV_0  \mcD^\da_{(t_1 - )} \mcV_0  \mcD_{ (t_2 -)} \mcV_0   \Delta^{--} (t_1, t_{2} ) 
}
\eeq
where we omitted the time arguments of the $\mcV_0$ operators.
The $T_F$ time ordering will now sort the operators in ascending order in time from right to left:
\beq
\es{
& i \mcV^{(1)} (t,0) =\\ =&   \int_0^t  dt_1 \int_0^{t_1} dt_2     \mcV_0  \mcD^\da_{(t_1 + )} \mcV_0  \mcD_{ (t_2 +)} \mcV_0   \Delta^{++} (t_1, t_{2} ) +\xi   \int_0^t  dt_1 \int_{t_1}^{t} dt_2     \mcV_0  \mcD_{ (t_2 +)} \mcV_0   \mcD^\da_{(t_1 + )} \mcV_0   \Delta^{++} (t_1, t_{2} ) + \\ 
 &-\xi  \int_0^t  dt_1 \int_0^{t_1} dt_2     \mcV_0  \mcD^\da_{(t_1+ )}  \mcV_0   \mcD_{ (t_2 -)}\mcV_0   \Delta^{+-} (t_1, t_{2} ) -  \xi \int_0^t  dt_1 \int_{t_1}^{t} dt_2     \mcV_0   \mcD_{ (t_2 -)}  \mcV_0  \mcD^\da_{(t_1 + )} \mcV_0   \Delta^{+-} (t_1, t_{2} ) + \\ 
 &-  \int_0^t  dt_1 \int_0^{t_1} dt_2     \mcV_0  \mcD^\da_{(t_1- )} \mcV_0  \mcD_{ (t_2+)} \mcV_0   \Delta^{-+} (t_1, t_{2} )  -  \int_0^t  dt_1 \int_{t_1}^{t} dt_2     \mcV_0  \mcD_{ (t_2 +)} \mcV_0   \mcD^\da_{(t_1 - )} \mcV_0   \Delta^{-+} (t_1, t_{2} ) + \\ 
&+\xi  \int_0^t  dt_1 \int_0^{t_1} dt_2     \mcV_0   \mcD^\da_{ (t_1 -)}  \mcV_0   \mcD_{(t_2- )}\mcV_0   \Delta^{--} (t_1, t_{2} )  +   \int_0^t  dt_1 \int_{t_1}^{t} dt_2     \mcV_0  \mcD_{ (t_2 -)}  \mcV_0  \mcD^\da_{(t_1 - )} \mcV_0   \Delta^{--} (t_1, t_{2} ) 
}
\eeq
Using the fact that $\int_0^{t} dt_1 \int_0^{t_1} dt_2 = \int_0^{t} dt_2 \int_0^{t_2} dt_1$ and exchanging the integration times in the second column we get
\beq
\es{
 i \mcV^{(1)} (t,0) &=\\ 
=   \int_0^t  dt_1 \int_0^{t_1} dt_2   [  &\mcV_0  \mcD^\da_{(t_1 + )} \mcV_0  \mcD_{ (t_2 +)} \mcV_0   \Delta^{++} (t_1, t_{2} ) +\xi \mcV_0  \mcD_{ (t_1 +)} \mcV_0   \mcD^\da_{(t_2 + )} \mcV_0   \Delta^{++} (t_2, t_{1} ) + \\ 
   -\xi &\mcV_0  \mcD^\da_{(t_1+ )}  \mcV_0   \mcD_{ (t_2 -)}\mcV_0   \Delta^{+-} (t_1, t_{2} ) -  \xi    \mcV_0   \mcD_{ (t_1 -)}  \mcV_0  \mcD^\da_{(t_2 + )} \mcV_0   \Delta^{+-} (t_2, t_{1} ) + \\ 
 - & \mcV_0  \mcD^\da_{(t_1- )} \mcV_0  \mcD_{ (t_2+)} \mcV_0   \Delta^{-+} (t_1, t_{2} )  -      \mcV_0  \mcD_{ (t_1 +)} \mcV_0   \mcD^\da_{(t_2 - )} \mcV_0   \Delta^{-+} (t_2, t_{1} ) + \\ 
 + \xi &\mcV_0   \mcD^\da_{ (t_1 -)}  \mcV_0   \mcD_{(t_2- )}\mcV_0   \Delta^{--} (t_1, t_{2} )  +     \mcV_0  \mcD_{ (t_1 -)}  \mcV_0  \mcD^\da_{(t_2 - )} \mcV_0   \Delta^{--} (t_2, t_{1} )  ] =
}
\eeq
\begin{equation*}
\es{
&=\sum_{\alpha \beta \in \{ + , - \} } \alpha^{(1+\xi)/2} \beta \int_0^t  dt_1 \int_0^{t_1} dt_2 \times  \\ 
\times  &\mcV_0(t,t_1) \lsq  \mcD^\da_{(t_1 \beta )} \mcV_0 (t_1,t_2)  \mcD_{ (t_2 \alpha)}   \Delta^{\beta \alpha} (t_1, t_{2} )  + \xi       \mcD_{ (t_1 \beta)} \mcV_0(t_1,t_2)   \mcD^\da_{(t_2 \alpha )}   \Delta^{\alpha\beta} (t_2, t_{1} )  \rsq \mcV_0(t_2,0) 
}
\end{equation*}
which defines the NCA self-energy
\begin{align}
\label{eq:NCASigma}
&\Sigma (t_1,t_2) =  \sum_{\alpha \beta \in \{ + , - \}  }   - \alpha^{(1+\xi)/2} \beta i \lsq    \mcD_{ \beta   }^\da \mcV (t_1,t_2)  \mcD_{ \alpha  } \Delta^{\beta  \alpha } (t_1, t_2 )  +\xi    \mcD_{\beta } \mcV (t_1,t_2)  \mcD^\da_{\alpha }  \Delta^{ \alpha   \beta } (t_2,t_1 )   \rsq 
\end{align}

\section{Trace preservation}

The evolution super-operator $\mcV$ obtained in the non-crossing approximation preserves the trace of the density operator, that is 
$$\tr \lbr \mcV(t,t')   \bullet  \rbr = \mcV(t,t')   \tr  \lbr \bullet \rbr $$
Writing $ \mcV(t,t')  \bullet $ means applying the super-operator $\mcV$ on a generic operator $\bullet$; we recall that in the usual representation where the operator $\bullet$ is a matrix, this is not a matrix product.
To prove this, we take the trace of the NCA Dyson equation and we use the fact that $\mcV_0$ does preserve the trace
\beq
\es{
\tr \lsq \mcV(t,t') \rho_I(t') \rsq &= \tr \lsq \mcV_0(t,t')\rho_I(t') \rsq + \int_{t'}^t dt_1 \int_{t'}^{t_1} dt_2 \tr \lsq \mcV_0(t,t_1) \Sigma(t_1,t_2) \mcV (t_2,t') \rho_I(t')  \rsq = \\ 
&= \tr \lsq \rho_I(t') \rsq + \int_{t'}^t dt_1 \int_{t'}^{t_1} dt_2 \tr \lsq  \Sigma(t_1,t_2)  \mcV (t_2,t')  \rho_I(t')  \rsq 
}
\eeq
Then, using the expression for the self-energy eq. \eqref{eq:NCASigma}, we can prove that the integrand vanishes, which completes the proof.
In order to show that, we remark that for the cyclic property of the trace it holds that $\tr \lsq  X_+ \bullet \rsq =\tr \lsq  X_- \bullet \rsq  $, $X$ being a generic super-operator. Then we can fix the $ \mcD^\da_\beta,  \mcD_\beta$ super-operators to be $ \mcD^\da_+,  \mcD_+$ under the trace, getting 
\beq
\es{
 &\tr \lsq  \Sigma(t_1,t_2)  \mcV (t_2,t')  \rho_I(t')  \rsq = \\ 
 = &\sum_{a , b} \sum_{\alpha \beta \in \{ + , - \}  }   - \alpha^{(1+\xi)/2} \beta \, i \,  \tr \lbr \lsq  \Delta_{b  a }^{\beta \alpha} (t_1, t_2 ) \mcD_{ \beta  b }^\da \mcV (t_1,t_2) \mcD_{ \alpha a } + \xi  \Delta_{ a b }^{\alpha \beta} (t_2,t_1 ) \mcD_{\beta b} \mcV (t_1,t_2) \mcD^\da_{\alpha a}   \rsq   \mcV (t_1,t') \rho_I(t') \rbr  = \\ 
  = &\sum_{a , b} \sum_{\alpha \beta \in \{ + , - \}  }   - \alpha^{(1+\xi)/2} \beta \, i \,  \tr \lbr \lsq  \Delta_{b  a }^{\beta \alpha} (t_1, t_2 ) \mcD_{ +  b }^\da \mcV (t_1,t_2) \mcD_{ \alpha a } + \xi  \Delta_{ a b }^{\alpha \beta} (t_2,t_1 ) \mcD_{+ b} \mcV (t_1,t_2) \mcD^\da_{\alpha a}   \rsq   \mcV (t_1,t') \rho_I(t') \rbr  
 }
 \eeq

By summing over $\beta$, one gets the two terms $ \Delta_{b  a }^{+ \alpha} (t_1, t_2 ) -\Delta_{b  a }^{- \alpha} (t_1, t_2 )  $ and $ \Delta_{ a b }^{\alpha +} (t_2,t_1 )-\Delta_{ a b }^{\alpha -} (t_2,t_1 )$, which vanish because of the following identities, holding for $t_1>t_2$: 
\begin{align*}
\Delta^{++}(t_1,t_2) = \Delta^{-+}(t_1,t_2)  &&
\Delta^{++}(t_2,t_1)= \Delta^{+-}(t_2,t_1)  \\ 
\Delta^{+-}(t_1,t_2) = \Delta^{--}(t_1,t_2)  && 
\Delta^{-+}(t_2,t_1)= \Delta^{--}(t_2,t_1)
\end{align*}
These identities hold because of the definition of $\Delta$, given in the main text, in terms of contour time-ordered Green's functions, i.e. $\Delta^{\alpha,\beta}( {t_1},{t_2}) \sim - i \aver{T_C c(t_1,\alpha) c^\da (t_2,\beta)} $, and remembering that times on the $-$ contour branch come after times on the $+$ one.
%

\subsection{Spectral properties the propagator $\mcV$}
{We call $\lambda_i(t,t'), v^R_i (t,t'),v^L_i (t,t')$ the eigenvalues and right and left eigenvectors of $\mcV (t,t')$, which depend on time.
As it preserves the trace, $\mcV (t,t')$ must have at least one eigenvalue equal to one, say $\lambda_0 \equiv 1$. 
If we assume this eigenvalue is non-degenerate, then all the others eigenvectors with $i \neq 0 $, are traceless. 
The proof of these properties goes as follows.}
$\tr{ \lsq  \mcV(t,t')\rho \rsq } = \tr \lsq \rho \rsq$ in the matrix notation reads $ \bra{\id} \bar{\bar{\mcV }} (t,t') \ket{\rho} =\aver{\id | \rho}$ which holds for every $\ket{\rho}$ {as the trace is preserved}; then $\bra{\id}$ must be a left eigenvector of $\mcV (t,t')$, $\bra{v^L_0 } \equiv \bra{\id } $, with eigenvalue $\lambda_0 = 1$. If we assume that there is only one eigenvector with eigenvalue 1, then all the others right-eigenvectors of $\mcV (t,t')$ must be orthogonal to $\bra{\id}$, that is they must have zero trace:  $\aver{\id | v_i^R(t,t') } = \tr{ \lsq v_i^R(t,t') \rsq} = 0$, for $i\neq 0$. 

\section{Hermiticity preservation}

A quantum dynamical map evolving the density operator should preserve its Hermiticity: will show that this property is not spoiled by the NCA approximation. 
The proof is inductive and goes as showing that, if $\mcV(t,t')$ preserves Hermiticity, then $\mcV(t+dt,t')$ does; given the initial condition $\mcV(t',t')=\id$, that obviously preserves Hermiticity, then it follows that $\mcV(t,t')$ is Hermiticity preserving $\forall t$.
Assuming $\mcV(t,t')$ is analytic in $t$, then its increment is given by its Taylor series
\beq
\mcV(t+dt ,t' ) = \mcV(t,t') + dt \, \pt \mcV(t,t') + \frac{dt^2}{2}\pt^2 \mcV(t,t') + \dots
\eeq

From the Dyson equation \eqref{eq:dyson}, we can show that if $\mcV(t,t')$ preserves Hermiticity, then all its derivatives do, which in turn implies, from the Taylor expansion, that $\mcV(t+dt ,t' ) $ does. This ultimately comes for the causal structure of the Dyson equation.

We will restrict to show that $\pt \mcV(t,t')$ is Hermiticity preserving, assuming $\mcV(t,t')$ is.
This result can be generalized to higher order derivatives, obtained by taking derivatives of the Dyson equation \eqref{eq:dyson}, with two observations: the n-th derivative of $\mcV$ depends only on its lower order derivatives; the structure of the equation for the n-th derivative is such that, if the lower order derivatives are Hermiticity preserving, then also the n-th derivative is. 

We now assume $\mcV(t,t')$ Hermiticity preserving and show that this implies $\pt \mcV(t,t')$ also is. With $\bullet $ an Hermitian operator, Hermiticity preservation of $\mcV$ means $\lp \mcV \bullet \rp^\da = \mcV \bullet $. $\mathcal{L}$ preserves Hermiticity as it is a Lindblad generator. Then, taking the hermitian conjugate of the Dyson equation \eqref{eq:dyson} we get

\beq 
\es{
\lp \pt \mcV(t,t') \bullet \rp^\da &= \lp  \mathcal{L} \mcV(t,t') \bullet  + \int_{t'}^t  dt_1 \Sigma(t,t_1) \mcV (t_1,t')  \bullet \rp^\da = \\ &=  \mathcal{L}\mcV(t,t') \bullet  + \int_{t'}^t  dt_1 \lp \Sigma(t,t_1) \mcV (t_1,t')  \bullet \rp^\da
}
\eeq

%
We need to determine the hermitian conjugate of $ \Sigma(t_1,t_2) \, \bullet $. $\Sigma(t_1,t_2) $ in the NCA approximation is given by \eqref{eq:NCASigma} and it depends on the hybridization function $\Delta$.
Defining the Keldysh indices $\bar{\alpha} = - \alpha$, $\bar{\beta} = - \beta $ and with $^*$ meaning complex conjugation, the following property holds $$ \lp \Delta^{\alpha \beta}  (t_1,t_2) \rp^* = - \Delta^{\bar{\beta} \bar{\alpha}}  (t_2,t_1)$$
that can be proven from the definition of $\Delta^{\alpha,\beta}( {t_1},{t_2}) \propto - i \aver{T_C c(t_1,\alpha) c^\da (t_2,\beta)} $ and writing down explicitly its Keldysh components.
It also holds that $$\lp X_\alpha \bullet \rp^\da = X_{\bar{\alpha}}^\da \, \bullet^\da  $$ as $\lp X_+ \bullet \rp^\da = \lp X \bullet \rp^\da =   \bullet^\da X^\da = X_-^\da \bullet^\da$ and  $\lp X_- \bullet \rp^\da = \lp \bullet X \rp^\da =   X^\da \bullet^\da  = X_+^\da \bullet^\da$.
For a nested application of super-operators as it appears in the self-energy, this property gives $\lp X_{1\alpha} \mcV X_{2 \beta } \bullet \rp^\da = X^\da_{1\bar{\alpha}} \mcV X^\da_{2 \bar{\beta} } \,  \bullet $, where we have used that the our ansaz for $\mcV$ preserves hermiticity and that $\bullet$ is hermitian.
Using these two results it follows that $\lp \Sigma (t_1,t_2) \bullet \rp^\da = \Sigma (t_1,t_2) \bullet  $ as 

\beq
\label{eq:NCASigmaHermiticity}
\es{
&\lp \Sigma (t_1,t_2) \bullet \rp^\da  =  \lp \sum_{\alpha \beta \in \{ + , - \}  }   - \alpha^{(1+\xi)/2} \beta i \lsq    \mcD_{ \beta   }^\da \mcV (t_1,t_2)  \mcD_{ \alpha  } \Delta^{\beta  \alpha } (t_1, t_2 )  +\xi    \mcD_{\beta } \mcV (t_1,t_2)  \mcD^\da_{\alpha }  \Delta^{ \alpha   \beta } (t_2,t_1 )   \rsq \bullet   \rp^\da = \\ 
&= \sum_{\alpha \beta \in \{ + , - \}  }   - \alpha^{(1+\xi)/2} \beta i \lsq    \mcD_{ \bar{\beta  } } \mcV (t_1,t_2)  \mcD_{ \bar{\alpha } }^\da \lp - \Delta^{ \bar{\alpha} \bar{\beta }} (t_2, t_1 )  \rp +\xi    \mcD^\da_{\bar{\beta} } \mcV (t_1,t_2)  \mcD_{\bar{\alpha} }  \lp - \Delta^{\bar{   \beta } \bar{ \alpha}} (t_1,t_2 )  \rp \rsq \bullet    = \\
&= \sum_{\bar{\alpha} \bar{\beta} \in \{ + , - \}  }   \xi \lp -1 \rp^{{(1+\xi)/2}} \bar{\alpha}^{(1+\xi)/2} \bar{\beta} i \lsq   \xi \mcD_{ \bar{\beta  } } \mcV (t_1,t_2)  \mcD_{ \bar{\alpha } }^\da\Delta^{ \bar{\alpha} \bar{\beta }} (t_2, t_1 ) +    \mcD^\da_{\bar{\beta} } \mcV (t_1,t_2)  \mcD_{\bar{\alpha} }  \Delta^{\bar{   \beta } \bar{ \alpha}} (t_1,t_2 )   \rsq \bullet    = \\ 
&= \Sigma (t_1,t_2) \bullet
}
\eeq
In the one but last equality, $\xi \lp -1 \rp^{{(1+\xi)/2}} = -1$, for both bosons and fermions $\xi=\pm 1$.
This completes the proof as 
\beq 
\es{
\lp \pt \mcV(t,t') \bullet \rp^\da &=  \mcV(t,t') \bullet  + \int_{t'}^t  dt_1 \lp \Sigma(t,t_1) \mcV (t_1,t')  \bullet \rp^\da =  \\ &= \mcV(t,t') \bullet  + \int_{t'}^t  dt_1  \Sigma(t,t_1) \mcV (t_1,t')  \bullet =  \pt \mcV(t,t') \bullet
}
\eeq

\begin{thebibliography}{58}%
\makeatletter
\providecommand \@ifxundefined [1]{%
 \@ifx{#1\undefined}
}%
\providecommand \@ifnum [1]{%
 \ifnum #1\expandafter \@firstoftwo
 \else \expandafter \@secondoftwo
 \fi
}%
\providecommand \@ifx [1]{%
 \ifx #1\expandafter \@firstoftwo
 \else \expandafter \@secondoftwo
 \fi
}%
\providecommand \natexlab [1]{#1}%
\providecommand \enquote  [1]{``#1''}%
\providecommand \bibnamefont  [1]{#1}%
\providecommand \bibfnamefont [1]{#1}%
\providecommand \citenamefont [1]{#1}%
\providecommand \href@noop [0]{\@secondoftwo}%
\providecommand \href [0]{\begingroup \@sanitize@url \@href}%
\providecommand \@href[1]{\@@startlink{#1}\@@href}%
\providecommand \@@href[1]{\endgroup#1\@@endlink}%
\providecommand \@sanitize@url [0]{\catcode `\\12\catcode `\$12\catcode
  `\&12\catcode `\#12\catcode `\^12\catcode `\_12\catcode `\%12\relax}%
\providecommand \@@startlink[1]{}%
\providecommand \@@endlink[0]{}%
\providecommand \url  [0]{\begingroup\@sanitize@url \@url }%
\providecommand \@url [1]{\endgroup\@href {#1}{\urlprefix }}%
\providecommand \urlprefix  [0]{URL }%
\providecommand \Eprint [0]{\href }%
\providecommand \doibase [0]{http://dx.doi.org/}%
\providecommand \selectlanguage [0]{\@gobble}%
\providecommand \bibinfo  [0]{\@secondoftwo}%
\providecommand \bibfield  [0]{\@secondoftwo}%
\providecommand \translation [1]{[#1]}%
\providecommand \BibitemOpen [0]{}%
\providecommand \bibitemStop [0]{}%
\providecommand \bibitemNoStop [0]{.\EOS\space}%
\providecommand \EOS [0]{\spacefactor3000\relax}%
\providecommand \BibitemShut  [1]{\csname bibitem#1\endcsname}%
\let\auto@bib@innerbib\@empty
\bibitem [{\citenamefont {Caldeira}\ and\ \citenamefont
  {Leggett}(1981)}]{Caldeira_Leggett_PRL81}%
  \BibitemOpen
  \bibfield  {author} {\bibinfo {author} {\bibfnamefont {A.~O.}\ \bibnamefont
  {Caldeira}}\ and\ \bibinfo {author} {\bibfnamefont {A.~J.}\ \bibnamefont
  {Leggett}},\ }\bibfield  {title} {\enquote {\bibinfo {title} {Influence of
  dissipation on quantum tunneling in macroscopic systems},}\ }\href {\doibase
  10.1103/PhysRevLett.46.211} {\bibfield  {journal} {\bibinfo  {journal} {Phys.
  Rev. Lett.}\ }\textbf {\bibinfo {volume} {46}},\ \bibinfo {pages} {211}
  (\bibinfo {year} {1981})}\BibitemShut {NoStop}%
\bibitem [{\citenamefont {Leggett}\ \emph {et~al.}(1987)\citenamefont
  {Leggett}, \citenamefont {Chakravarty}, \citenamefont {Dorsey}, \citenamefont
  {Fisher}, \citenamefont {Garg},\ and\ \citenamefont {Zwerger}}]{Leggett-RMP}%
  \BibitemOpen
  \bibfield  {author} {\bibinfo {author} {\bibfnamefont {A.~J.}\ \bibnamefont
  {Leggett}}, \bibinfo {author} {\bibfnamefont {S.}~\bibnamefont
  {Chakravarty}}, \bibinfo {author} {\bibfnamefont {A.~T.}\ \bibnamefont
  {Dorsey}}, \bibinfo {author} {\bibfnamefont {M.~P.~A.}\ \bibnamefont
  {Fisher}}, \bibinfo {author} {\bibfnamefont {A.}~\bibnamefont {Garg}}, \ and\
  \bibinfo {author} {\bibfnamefont {W.}~\bibnamefont {Zwerger}},\ }\bibfield
  {title} {\enquote {\bibinfo {title} {Dynamics of the dissipative two-state
  system},}\ }\href {\doibase 10.1103/RevModPhys.59.1} {\bibfield  {journal}
  {\bibinfo  {journal} {Rev. Mod. Phys.}\ }\textbf {\bibinfo {volume} {59}},\
  \bibinfo {pages} {1} (\bibinfo {year} {1987})}\BibitemShut {NoStop}%
\bibitem [{\citenamefont {Hur}\ \emph {et~al.}(2018)\citenamefont {Hur},
  \citenamefont {Henriet}, \citenamefont {Herviou}, \citenamefont {Plekhanov},
  \citenamefont {Petrescu}, \citenamefont {Goren}, \citenamefont {Schiro},
  \citenamefont {Mora},\ and\ \citenamefont {Orth}}]{LEHUR2018451}%
  \BibitemOpen
  \bibfield  {author} {\bibinfo {author} {\bibfnamefont {K.~L.}\ \bibnamefont
  {Hur}}, \bibinfo {author} {\bibfnamefont {L.}~\bibnamefont {Henriet}},
  \bibinfo {author} {\bibfnamefont {L.}~\bibnamefont {Herviou}}, \bibinfo
  {author} {\bibfnamefont {K.}~\bibnamefont {Plekhanov}}, \bibinfo {author}
  {\bibfnamefont {A.}~\bibnamefont {Petrescu}}, \bibinfo {author}
  {\bibfnamefont {T.}~\bibnamefont {Goren}}, \bibinfo {author} {\bibfnamefont
  {M.}~\bibnamefont {Schiro}}, \bibinfo {author} {\bibfnamefont
  {C.}~\bibnamefont {Mora}}, \ and\ \bibinfo {author} {\bibfnamefont {P.~P.}\
  \bibnamefont {Orth}},\ }\bibfield  {title} {\enquote {\bibinfo {title}
  {Driven dissipative dynamics and topology of quantum impurity systems},}\
  }\href {\doibase https://doi.org/10.1016/j.crhy.2018.04.003} {\bibfield
  {journal} {\bibinfo  {journal} {Comptes Rendus Physique}\ }\textbf {\bibinfo
  {volume} {19}},\ \bibinfo {pages} {451 -- 483} (\bibinfo {year} {2018})},\
  \bibinfo {note} {quantum simulation / Simulation quantique}\BibitemShut
  {NoStop}%
\bibitem [{\citenamefont {Hewson}(1993)}]{Hewson_book}%
  \BibitemOpen
  \bibfield  {author} {\bibinfo {author} {\bibfnamefont {A.~C.}\ \bibnamefont
  {Hewson}},\ }\href@noop {} {\emph {\bibinfo {title} {The Kondo Problem to
  Heavy Fermions}}}\ (\bibinfo  {publisher} {Cambridge University Press},\
  \bibinfo {year} {1993})\BibitemShut {NoStop}%
\bibitem [{\citenamefont {Goldhaber-Gordon~\textit{et
  al.}}(1998)}]{Goldhaber_Gordon_nature98}%
  \BibitemOpen
  \bibfield  {author} {\bibinfo {author} {\bibfnamefont {D.}~\bibnamefont
  {Goldhaber-Gordon~\textit{et al.}}},\ }\bibfield  {title} {\enquote {\bibinfo
  {title} {{Kondo effect in a single-electron transistor}},}\ }\href@noop {}
  {\bibfield  {journal} {\bibinfo  {journal} {Nature}\ }\textbf {\bibinfo
  {volume} {391}},\ \bibinfo {pages} {156} (\bibinfo {year}
  {1998})}\BibitemShut {NoStop}%
\bibitem [{\citenamefont {Grobis}\ \emph {et~al.}(2008)\citenamefont {Grobis},
  \citenamefont {Rau}, \citenamefont {Potok}, \citenamefont {Shtrikman},\ and\
  \citenamefont {Goldhaber-Gordon}}]{GoldhaberGordon_prl08_noneq}%
  \BibitemOpen
  \bibfield  {author} {\bibinfo {author} {\bibfnamefont {M.}~\bibnamefont
  {Grobis}}, \bibinfo {author} {\bibfnamefont {I.~G.}\ \bibnamefont {Rau}},
  \bibinfo {author} {\bibfnamefont {R.~M.}\ \bibnamefont {Potok}}, \bibinfo
  {author} {\bibfnamefont {H.}~\bibnamefont {Shtrikman}}, \ and\ \bibinfo
  {author} {\bibfnamefont {D.}~\bibnamefont {Goldhaber-Gordon}},\ }\bibfield
  {title} {\enquote {\bibinfo {title} {{Universal Scaling in Nonequilibrium
  Transport through a Single Channel Kondo Dot}},}\ }\href {\doibase
  10.1103/PhysRevLett.100.246601} {\bibfield  {journal} {\bibinfo  {journal}
  {Phys. Rev. Lett.}\ }\textbf {\bibinfo {volume} {100}},\ \bibinfo {pages}
  {246601} (\bibinfo {year} {2008})}\BibitemShut {NoStop}%
\bibitem [{\citenamefont {Roch}\ \emph {et~al.}(2008)\citenamefont {Roch},
  \citenamefont {Florens}, \citenamefont {Bouchiat}, \citenamefont
  {Wernsdorfer},\ and\ \citenamefont {Balestro}}]{Florens_C60}%
  \BibitemOpen
  \bibfield  {author} {\bibinfo {author} {\bibfnamefont {N.}~\bibnamefont
  {Roch}}, \bibinfo {author} {\bibfnamefont {S.}~\bibnamefont {Florens}},
  \bibinfo {author} {\bibfnamefont {V.}~\bibnamefont {Bouchiat}}, \bibinfo
  {author} {\bibfnamefont {W.}~\bibnamefont {Wernsdorfer}}, \ and\ \bibinfo
  {author} {\bibfnamefont {F.}~\bibnamefont {Balestro}},\ }\bibfield  {title}
  {\enquote {\bibinfo {title} {Quantum phase transition in a single-molecule
  quantum dot},}\ }\href@noop {} {\bibfield  {journal} {\bibinfo  {journal}
  {Nature}\ }\textbf {\bibinfo {volume} {453}} (\bibinfo {year}
  {2008})}\BibitemShut {NoStop}%
\bibitem [{\citenamefont {Iftikhar}\ \emph {et~al.}(2018)\citenamefont
  {Iftikhar}, \citenamefont {Anthore}, \citenamefont {Mitchell}, \citenamefont
  {Parmentier}, \citenamefont {Gennser}, \citenamefont {Ouerghi}, \citenamefont
  {Cavanna}, \citenamefont {Mora}, \citenamefont {Simon},\ and\ \citenamefont
  {Pierre}}]{IftikharEtAlScience2018}%
  \BibitemOpen
  \bibfield  {author} {\bibinfo {author} {\bibfnamefont {Z.}~\bibnamefont
  {Iftikhar}}, \bibinfo {author} {\bibfnamefont {A.}~\bibnamefont {Anthore}},
  \bibinfo {author} {\bibfnamefont {A.~K.}\ \bibnamefont {Mitchell}}, \bibinfo
  {author} {\bibfnamefont {F.~D.}\ \bibnamefont {Parmentier}}, \bibinfo
  {author} {\bibfnamefont {U.}~\bibnamefont {Gennser}}, \bibinfo {author}
  {\bibfnamefont {A.}~\bibnamefont {Ouerghi}}, \bibinfo {author} {\bibfnamefont
  {A.}~\bibnamefont {Cavanna}}, \bibinfo {author} {\bibfnamefont
  {C.}~\bibnamefont {Mora}}, \bibinfo {author} {\bibfnamefont {P.}~\bibnamefont
  {Simon}}, \ and\ \bibinfo {author} {\bibfnamefont {F.}~\bibnamefont
  {Pierre}},\ }\bibfield  {title} {\enquote {\bibinfo {title} {Tunable quantum
  criticality and super-ballistic transport in a
  {\textquotedblleft}charge{\textquotedblright} kondo circuit},}\ }\href
  {\doibase 10.1126/science.aan5592} {\bibfield  {journal} {\bibinfo  {journal}
  {Science}\ }\textbf {\bibinfo {volume} {360}},\ \bibinfo {pages} {1315--1320}
  (\bibinfo {year} {2018})},\ \Eprint
  {http://arxiv.org/abs/http://science.sciencemag.org/content/360/6395/1315.full.pdf}
  {http://science.sciencemag.org/content/360/6395/1315.full.pdf} \BibitemShut
  {NoStop}%
\bibitem [{\citenamefont {Cohen}\ and\ \citenamefont
  {Rabani}(2011)}]{CohenRabaniPRB11}%
  \BibitemOpen
  \bibfield  {author} {\bibinfo {author} {\bibfnamefont {G.}~\bibnamefont
  {Cohen}}\ and\ \bibinfo {author} {\bibfnamefont {E.}~\bibnamefont {Rabani}},\
  }\bibfield  {title} {\enquote {\bibinfo {title} {Memory effects in
  nonequilibrium quantum impurity models},}\ }\href {\doibase
  10.1103/PhysRevB.84.075150} {\bibfield  {journal} {\bibinfo  {journal} {Phys.
  Rev. B}\ }\textbf {\bibinfo {volume} {84}},\ \bibinfo {pages} {075150}
  (\bibinfo {year} {2011})}\BibitemShut {NoStop}%
\bibitem [{\citenamefont {Gull}\ \emph {et~al.}(2011)\citenamefont {Gull},
  \citenamefont {Millis}, \citenamefont {Lichtenstein}, \citenamefont
  {Rubtsov}, \citenamefont {Troyer},\ and\ \citenamefont
  {Werner}}]{Gull_RMP11}%
  \BibitemOpen
  \bibfield  {author} {\bibinfo {author} {\bibfnamefont {E.}~\bibnamefont
  {Gull}}, \bibinfo {author} {\bibfnamefont {A.~J.}\ \bibnamefont {Millis}},
  \bibinfo {author} {\bibfnamefont {A.~I.}\ \bibnamefont {Lichtenstein}},
  \bibinfo {author} {\bibfnamefont {A.~N.}\ \bibnamefont {Rubtsov}}, \bibinfo
  {author} {\bibfnamefont {M.}~\bibnamefont {Troyer}}, \ and\ \bibinfo {author}
  {\bibfnamefont {P.}~\bibnamefont {Werner}},\ }\bibfield  {title} {\enquote
  {\bibinfo {title} {Continuous-time monte~carlo methods for quantum impurity
  models},}\ }\href {\doibase 10.1103/RevModPhys.83.349} {\bibfield  {journal}
  {\bibinfo  {journal} {Rev. Mod. Phys.}\ }\textbf {\bibinfo {volume} {83}},\
  \bibinfo {pages} {349--404} (\bibinfo {year} {2011})}\BibitemShut {NoStop}%
\bibitem [{\citenamefont {M\"uhlbacher}\ and\ \citenamefont
  {Rabani}(2008)}]{muhlbacherRabaniPRL2008}%
  \BibitemOpen
  \bibfield  {author} {\bibinfo {author} {\bibfnamefont {L.}~\bibnamefont
  {M\"uhlbacher}}\ and\ \bibinfo {author} {\bibfnamefont {E.}~\bibnamefont
  {Rabani}},\ }\bibfield  {title} {\enquote {\bibinfo {title} {Real-time path
  integral approach to nonequilibrium many-body quantum systems},}\ }\href
  {\doibase 10.1103/PhysRevLett.100.176403} {\bibfield  {journal} {\bibinfo
  {journal} {Phys. Rev. Lett.}\ }\textbf {\bibinfo {volume} {100}},\ \bibinfo
  {pages} {176403} (\bibinfo {year} {2008})}\BibitemShut {NoStop}%
\bibitem [{\citenamefont {Schir\'{o}}\ and\ \citenamefont
  {Fabrizio}(2009)}]{Keldysh_short}%
  \BibitemOpen
  \bibfield  {author} {\bibinfo {author} {\bibfnamefont {M.}~\bibnamefont
  {Schir\'{o}}}\ and\ \bibinfo {author} {\bibfnamefont {M.}~\bibnamefont
  {Fabrizio}},\ }\bibfield  {title} {\enquote {\bibinfo {title} {Real-time
  diagrammatic monte carlo for nonequilibrium quantum transport},}\ }\href
  {\doibase 10.1103/PhysRevB.79.153302} {\bibfield  {journal} {\bibinfo
  {journal} {Phys. Rev. B}\ }\textbf {\bibinfo {volume} {79}},\ \bibinfo
  {pages} {153302} (\bibinfo {year} {2009})}\BibitemShut {NoStop}%
\bibitem [{\citenamefont {Werner}, \citenamefont {Oka},\ and\ \citenamefont
  {Millis}(2009)}]{Werner_Keldysh_09}%
  \BibitemOpen
  \bibfield  {author} {\bibinfo {author} {\bibfnamefont {P.}~\bibnamefont
  {Werner}}, \bibinfo {author} {\bibfnamefont {T.}~\bibnamefont {Oka}}, \ and\
  \bibinfo {author} {\bibfnamefont {A.~J.}\ \bibnamefont {Millis}},\ }\bibfield
   {title} {\enquote {\bibinfo {title} {Diagrammatic monte carlo simulation of
  nonequilibrium systems},}\ }\href {\doibase 10.1103/PhysRevB.79.035320}
  {\bibfield  {journal} {\bibinfo  {journal} {Phys. Rev. B}\ }\textbf {\bibinfo
  {volume} {79}},\ \bibinfo {pages} {035320} (\bibinfo {year}
  {2009})}\BibitemShut {NoStop}%
\bibitem [{\citenamefont {Cohen}\ \emph {et~al.}(2013)\citenamefont {Cohen},
  \citenamefont {Gull}, \citenamefont {Reichman}, \citenamefont {Millis},\ and\
  \citenamefont {Rabani}}]{CohenEtAlPR13}%
  \BibitemOpen
  \bibfield  {author} {\bibinfo {author} {\bibfnamefont {G.}~\bibnamefont
  {Cohen}}, \bibinfo {author} {\bibfnamefont {E.}~\bibnamefont {Gull}},
  \bibinfo {author} {\bibfnamefont {D.~R.}\ \bibnamefont {Reichman}}, \bibinfo
  {author} {\bibfnamefont {A.~J.}\ \bibnamefont {Millis}}, \ and\ \bibinfo
  {author} {\bibfnamefont {E.}~\bibnamefont {Rabani}},\ }\bibfield  {title}
  {\enquote {\bibinfo {title} {Numerically exact long-time magnetization
  dynamics at the nonequilibrium kondo crossover of the anderson impurity
  model},}\ }\href {\doibase 10.1103/PhysRevB.87.195108} {\bibfield  {journal}
  {\bibinfo  {journal} {Phys. Rev. B}\ }\textbf {\bibinfo {volume} {87}},\
  \bibinfo {pages} {195108} (\bibinfo {year} {2013})}\BibitemShut {NoStop}%
\bibitem [{\citenamefont {Profumo}\ \emph {et~al.}(2015)\citenamefont
  {Profumo}, \citenamefont {Groth}, \citenamefont {Messio}, \citenamefont
  {Parcollet},\ and\ \citenamefont {Waintal}}]{ProfumoEtAlPRB15}%
  \BibitemOpen
  \bibfield  {author} {\bibinfo {author} {\bibfnamefont {R.~E.~V.}\
  \bibnamefont {Profumo}}, \bibinfo {author} {\bibfnamefont {C.}~\bibnamefont
  {Groth}}, \bibinfo {author} {\bibfnamefont {L.}~\bibnamefont {Messio}},
  \bibinfo {author} {\bibfnamefont {O.}~\bibnamefont {Parcollet}}, \ and\
  \bibinfo {author} {\bibfnamefont {X.}~\bibnamefont {Waintal}},\ }\bibfield
  {title} {\enquote {\bibinfo {title} {Quantum monte carlo for correlated
  out-of-equilibrium nanoelectronic devices},}\ }\href@noop {} {\bibfield
  {journal} {\bibinfo  {journal} {Phys. Rev. B}\ }\textbf {\bibinfo {volume}
  {91}},\ \bibinfo {pages} {245154} (\bibinfo {year} {2015})}\BibitemShut
  {NoStop}%
\bibitem [{\citenamefont {Cohen}\ \emph {et~al.}(2015)\citenamefont {Cohen},
  \citenamefont {Gull}, \citenamefont {Reichman},\ and\ \citenamefont
  {Millis}}]{CohenEtAlPRL15}%
  \BibitemOpen
  \bibfield  {author} {\bibinfo {author} {\bibfnamefont {G.}~\bibnamefont
  {Cohen}}, \bibinfo {author} {\bibfnamefont {E.}~\bibnamefont {Gull}},
  \bibinfo {author} {\bibfnamefont {D.~R.}\ \bibnamefont {Reichman}}, \ and\
  \bibinfo {author} {\bibfnamefont {A.~J.}\ \bibnamefont {Millis}},\ }\bibfield
   {title} {\enquote {\bibinfo {title} {Taming the dynamical sign problem in
  real-time evolution of quantum many-body problems},}\ }\href@noop {}
  {\bibfield  {journal} {\bibinfo  {journal} {Phys. Rev. Lett.}\ }\textbf
  {\bibinfo {volume} {115}},\ \bibinfo {pages} {266802} (\bibinfo {year}
  {2015})}\BibitemShut {NoStop}%
\bibitem [{\citenamefont {Chen}, \citenamefont {Cohen},\ and\ \citenamefont
  {Reichman}(2017{\natexlab{a}})}]{ThetaJCP17_1}%
  \BibitemOpen
  \bibfield  {author} {\bibinfo {author} {\bibfnamefont {H.-T.}\ \bibnamefont
  {Chen}}, \bibinfo {author} {\bibfnamefont {G.}~\bibnamefont {Cohen}}, \ and\
  \bibinfo {author} {\bibfnamefont {D.~R.}\ \bibnamefont {Reichman}},\
  }\bibfield  {title} {\enquote {\bibinfo {title} {Inchworm monte carlo for
  exact non-adiabatic dynamics. i. theory and algorithms},}\ }\href {\doibase
  10.1063/1.4974328} {\bibfield  {journal} {\bibinfo  {journal} {The Journal of
  Chemical Physics}\ }\textbf {\bibinfo {volume} {146}},\ \bibinfo {pages}
  {054105} (\bibinfo {year} {2017}{\natexlab{a}})},\ \Eprint
  {http://arxiv.org/abs/https://doi.org/10.1063/1.4974328}
  {https://doi.org/10.1063/1.4974328} \BibitemShut {NoStop}%
\bibitem [{\citenamefont {Chen}, \citenamefont {Cohen},\ and\ \citenamefont
  {Reichman}(2017{\natexlab{b}})}]{ThetaJCP17_2}%
  \BibitemOpen
  \bibfield  {author} {\bibinfo {author} {\bibfnamefont {H.-T.}\ \bibnamefont
  {Chen}}, \bibinfo {author} {\bibfnamefont {G.}~\bibnamefont {Cohen}}, \ and\
  \bibinfo {author} {\bibfnamefont {D.~R.}\ \bibnamefont {Reichman}},\
  }\bibfield  {title} {\enquote {\bibinfo {title} {Inchworm monte carlo for
  exact non-adiabatic dynamics. ii. benchmarks and comparison with established
  methods},}\ }\href {\doibase 10.1063/1.4974329} {\bibfield  {journal}
  {\bibinfo  {journal} {The Journal of Chemical Physics}\ }\textbf {\bibinfo
  {volume} {146}},\ \bibinfo {pages} {054106} (\bibinfo {year}
  {2017}{\natexlab{b}})},\ \Eprint
  {http://arxiv.org/abs/https://doi.org/10.1063/1.4974329}
  {https://doi.org/10.1063/1.4974329} \BibitemShut {NoStop}%
\bibitem [{\citenamefont {Bloch}, \citenamefont {Dalibard},\ and\ \citenamefont
  {Nascimb{\`e}ne}(2012)}]{BlochDalibardNascimbeneNatPhys12}%
  \BibitemOpen
  \bibfield  {author} {\bibinfo {author} {\bibfnamefont {I.}~\bibnamefont
  {Bloch}}, \bibinfo {author} {\bibfnamefont {J.}~\bibnamefont {Dalibard}}, \
  and\ \bibinfo {author} {\bibfnamefont {S.}~\bibnamefont {Nascimb{\`e}ne}},\
  }\bibfield  {title} {\enquote {\bibinfo {title} {Quantum simulations with
  ultracold quantum gases},}\ }\href {http://dx.doi.org/10.1038/nphys2259}
  {\bibfield  {journal} {\bibinfo  {journal} {Nature Physics}\ }\textbf
  {\bibinfo {volume} {8}},\ \bibinfo {pages} {267 EP --} (\bibinfo {year}
  {2012})}\BibitemShut {NoStop}%
\bibitem [{\citenamefont {Blatt}\ and\ \citenamefont
  {Roos}(2012)}]{BlattRoosNatPhys12}%
  \BibitemOpen
  \bibfield  {author} {\bibinfo {author} {\bibfnamefont {R.}~\bibnamefont
  {Blatt}}\ and\ \bibinfo {author} {\bibfnamefont {C.~F.}\ \bibnamefont
  {Roos}},\ }\bibfield  {title} {\enquote {\bibinfo {title} {Quantum
  simulations with trapped ions},}\ }\href
  {http://dx.doi.org/10.1038/nphys2252} {\bibfield  {journal} {\bibinfo
  {journal} {Nature Physics}\ }\textbf {\bibinfo {volume} {8}},\ \bibinfo
  {pages} {277 EP --} (\bibinfo {year} {2012})}\BibitemShut {NoStop}%
\bibitem [{\citenamefont {Houck}, \citenamefont {Tureci},\ and\ \citenamefont
  {Koch}(2012)}]{AndrewNatPhys}%
  \BibitemOpen
  \bibfield  {author} {\bibinfo {author} {\bibfnamefont {A.~A.}\ \bibnamefont
  {Houck}}, \bibinfo {author} {\bibfnamefont {H.~E.}\ \bibnamefont {Tureci}}, \
  and\ \bibinfo {author} {\bibfnamefont {J.}~\bibnamefont {Koch}},\ }\bibfield
  {title} {\enquote {\bibinfo {title} {On-chip quantum simulation with
  superconducting circuits},}\ }\href@noop {} {\bibfield  {journal} {\bibinfo
  {journal} {Nature Physics}\ }\textbf {\bibinfo {volume} {8}} (\bibinfo {year}
  {2012})}\BibitemShut {NoStop}%
\bibitem [{\citenamefont {Hur}\ \emph {et~al.}(2016)\citenamefont {Hur},
  \citenamefont {Henriet}, \citenamefont {Petrescu}, \citenamefont {Plekhanov},
  \citenamefont {Roux},\ and\ \citenamefont {Schir\'o}}]{LeHurReview16}%
  \BibitemOpen
  \bibfield  {author} {\bibinfo {author} {\bibfnamefont {K.~L.}\ \bibnamefont
  {Hur}}, \bibinfo {author} {\bibfnamefont {L.}~\bibnamefont {Henriet}},
  \bibinfo {author} {\bibfnamefont {A.}~\bibnamefont {Petrescu}}, \bibinfo
  {author} {\bibfnamefont {K.}~\bibnamefont {Plekhanov}}, \bibinfo {author}
  {\bibfnamefont {G.}~\bibnamefont {Roux}}, \ and\ \bibinfo {author}
  {\bibfnamefont {M.}~\bibnamefont {Schir\'o}},\ }\bibfield  {title} {\enquote
  {\bibinfo {title} {Many-body quantum electrodynamics networks:
  Non-equilibrium condensed matter physics with light},}\ }\href {\doibase
  http://dx.doi.org/10.1016/j.crhy.2016.05.003} {\bibfield  {journal} {\bibinfo
   {journal} {Comptes Rendus Physique}\ }\textbf {\bibinfo {volume} {17}},\
  \bibinfo {pages} {808 -- 835} (\bibinfo {year} {2016})},\ \bibinfo {note}
  {polariton physics / Physique des polaritons}\BibitemShut {NoStop}%
\bibitem [{\citenamefont {Breuer}\ and\ \citenamefont
  {Petruccione}(2002)}]{BreuerPetruccione}%
  \BibitemOpen
  \bibfield  {author} {\bibinfo {author} {\bibfnamefont {H.-P.}\ \bibnamefont
  {Breuer}}\ and\ \bibinfo {author} {\bibfnamefont {F.}~\bibnamefont
  {Petruccione}},\ }\href@noop {} {\emph {\bibinfo {title} {The theory of open
  quantum systems}}},\ \bibinfo {edition} {1st}\ ed.\ (\bibinfo  {publisher}
  {Oxford University Press, {USA}},\ \bibinfo {year} {2002})\BibitemShut
  {NoStop}%
\bibitem [{\citenamefont {Diehl}\ \emph {et~al.}(2008)\citenamefont {Diehl},
  \citenamefont {Micheli}, \citenamefont {Kantian}, \citenamefont {Kraus},
  \citenamefont {Buchler},\ and\ \citenamefont {Zoller}}]{DiehlEtalNatPhys08}%
  \BibitemOpen
  \bibfield  {author} {\bibinfo {author} {\bibfnamefont {S.}~\bibnamefont
  {Diehl}}, \bibinfo {author} {\bibfnamefont {A.}~\bibnamefont {Micheli}},
  \bibinfo {author} {\bibfnamefont {A.}~\bibnamefont {Kantian}}, \bibinfo
  {author} {\bibfnamefont {B.}~\bibnamefont {Kraus}}, \bibinfo {author}
  {\bibfnamefont {H.~P.}\ \bibnamefont {Buchler}}, \ and\ \bibinfo {author}
  {\bibfnamefont {P.}~\bibnamefont {Zoller}},\ }\bibfield  {title} {\enquote
  {\bibinfo {title} {Quantum states and phases in driven open quantum systems
  with cold atoms},}\ }\href {http://dx.doi.org/10.1038/nphys1073} {\bibfield
  {journal} {\bibinfo  {journal} {Nat Phys}\ }\textbf {\bibinfo {volume} {4}},\
  \bibinfo {pages} {878--883} (\bibinfo {year} {2008})}\BibitemShut {NoStop}%
\bibitem [{\citenamefont {Verstraete}, \citenamefont {Wolf},\ and\
  \citenamefont {Ignacio~Cirac}(2009)}]{VerstraeteWolfCiracNatPhys09}%
  \BibitemOpen
  \bibfield  {author} {\bibinfo {author} {\bibfnamefont {F.}~\bibnamefont
  {Verstraete}}, \bibinfo {author} {\bibfnamefont {M.~M.}\ \bibnamefont
  {Wolf}}, \ and\ \bibinfo {author} {\bibfnamefont {J.}~\bibnamefont
  {Ignacio~Cirac}},\ }\bibfield  {title} {\enquote {\bibinfo {title} {Quantum
  computation and quantum-state engineering driven by dissipation},}\
  }\href@noop {} {\bibfield  {journal} {\bibinfo  {journal} {Nat Phys}\
  }\textbf {\bibinfo {volume} {5}},\ \bibinfo {pages} {633--636} (\bibinfo
  {year} {2009})}\BibitemShut {NoStop}%
\bibitem [{\citenamefont {Murch}\ \emph {et~al.}(2012)\citenamefont {Murch},
  \citenamefont {Vool}, \citenamefont {Zhou}, \citenamefont {Weber},
  \citenamefont {Girvin},\ and\ \citenamefont
  {Siddiqi}}]{Siddiqi_quantum_bath_engineering}%
  \BibitemOpen
  \bibfield  {author} {\bibinfo {author} {\bibfnamefont {K.~W.}\ \bibnamefont
  {Murch}}, \bibinfo {author} {\bibfnamefont {U.}~\bibnamefont {Vool}},
  \bibinfo {author} {\bibfnamefont {D.}~\bibnamefont {Zhou}}, \bibinfo {author}
  {\bibfnamefont {S.~J.}\ \bibnamefont {Weber}}, \bibinfo {author}
  {\bibfnamefont {S.~M.}\ \bibnamefont {Girvin}}, \ and\ \bibinfo {author}
  {\bibfnamefont {I.}~\bibnamefont {Siddiqi}},\ }\bibfield  {title} {\enquote
  {\bibinfo {title} {Cavity-assisted quantum bath engineering},}\ }\href@noop
  {} {\bibfield  {journal} {\bibinfo  {journal} {Phys. Rev. Lett.}\ }\textbf
  {\bibinfo {volume} {109}},\ \bibinfo {pages} {183602} (\bibinfo {year}
  {2012})}\BibitemShut {NoStop}%
\bibitem [{\citenamefont {Nakagawa}, \citenamefont {Kawakami},\ and\
  \citenamefont {Ueda}(2018)}]{NakagawaEtAlPRL18}%
  \BibitemOpen
  \bibfield  {author} {\bibinfo {author} {\bibfnamefont {M.}~\bibnamefont
  {Nakagawa}}, \bibinfo {author} {\bibfnamefont {N.}~\bibnamefont {Kawakami}},
  \ and\ \bibinfo {author} {\bibfnamefont {M.}~\bibnamefont {Ueda}},\
  }\bibfield  {title} {\enquote {\bibinfo {title} {Non-hermitian kondo effect
  in ultracold alkaline-earth atoms},}\ }\href {\doibase
  10.1103/PhysRevLett.121.203001} {\bibfield  {journal} {\bibinfo  {journal}
  {Phys. Rev. Lett.}\ }\textbf {\bibinfo {volume} {121}},\ \bibinfo {pages}
  {203001} (\bibinfo {year} {2018})}\BibitemShut {NoStop}%
\bibitem [{\citenamefont {Tonielli}\ \emph {et~al.}(2019)\citenamefont
  {Tonielli}, \citenamefont {Fazio}, \citenamefont {Diehl},\ and\ \citenamefont
  {Marino}}]{TonielliEtAlPRL19}%
  \BibitemOpen
  \bibfield  {author} {\bibinfo {author} {\bibfnamefont {F.}~\bibnamefont
  {Tonielli}}, \bibinfo {author} {\bibfnamefont {R.}~\bibnamefont {Fazio}},
  \bibinfo {author} {\bibfnamefont {S.}~\bibnamefont {Diehl}}, \ and\ \bibinfo
  {author} {\bibfnamefont {J.}~\bibnamefont {Marino}},\ }\bibfield  {title}
  {\enquote {\bibinfo {title} {Orthogonality catastrophe in dissipative quantum
  many-body systems},}\ }\href {\doibase 10.1103/PhysRevLett.122.040604}
  {\bibfield  {journal} {\bibinfo  {journal} {Phys. Rev. Lett.}\ }\textbf
  {\bibinfo {volume} {122}},\ \bibinfo {pages} {040604} (\bibinfo {year}
  {2019})}\BibitemShut {NoStop}%
\bibitem [{\citenamefont {Fr\"oml}\ \emph {et~al.}(2019)\citenamefont
  {Fr\"oml}, \citenamefont {Chiocchetta}, \citenamefont {Kollath},\ and\
  \citenamefont {Diehl}}]{HeinrichEtAlPRL19}%
  \BibitemOpen
  \bibfield  {author} {\bibinfo {author} {\bibfnamefont {H.}~\bibnamefont
  {Fr\"oml}}, \bibinfo {author} {\bibfnamefont {A.}~\bibnamefont
  {Chiocchetta}}, \bibinfo {author} {\bibfnamefont {C.}~\bibnamefont
  {Kollath}}, \ and\ \bibinfo {author} {\bibfnamefont {S.}~\bibnamefont
  {Diehl}},\ }\bibfield  {title} {\enquote {\bibinfo {title}
  {Fluctuation-induced quantum zeno effect},}\ }\href {\doibase
  10.1103/PhysRevLett.122.040402} {\bibfield  {journal} {\bibinfo  {journal}
  {Phys. Rev. Lett.}\ }\textbf {\bibinfo {volume} {122}},\ \bibinfo {pages}
  {040402} (\bibinfo {year} {2019})}\BibitemShut {NoStop}%
\bibitem [{\citenamefont {Delbecq}\ \emph {et~al.}(2011)\citenamefont
  {Delbecq}, \citenamefont {Schmitt}, \citenamefont {Parmentier}, \citenamefont
  {Roch}, \citenamefont {Viennot}, \citenamefont {F\`eve}, \citenamefont
  {Huard}, \citenamefont {Mora}, \citenamefont {Cottet},\ and\ \citenamefont
  {Kontos}}]{Delbecq_prl11}%
  \BibitemOpen
  \bibfield  {author} {\bibinfo {author} {\bibfnamefont {M.~R.}\ \bibnamefont
  {Delbecq}}, \bibinfo {author} {\bibfnamefont {V.}~\bibnamefont {Schmitt}},
  \bibinfo {author} {\bibfnamefont {F.~D.}\ \bibnamefont {Parmentier}},
  \bibinfo {author} {\bibfnamefont {N.}~\bibnamefont {Roch}}, \bibinfo {author}
  {\bibfnamefont {J.~J.}\ \bibnamefont {Viennot}}, \bibinfo {author}
  {\bibfnamefont {G.}~\bibnamefont {F\`eve}}, \bibinfo {author} {\bibfnamefont
  {B.}~\bibnamefont {Huard}}, \bibinfo {author} {\bibfnamefont
  {C.}~\bibnamefont {Mora}}, \bibinfo {author} {\bibfnamefont {A.}~\bibnamefont
  {Cottet}}, \ and\ \bibinfo {author} {\bibfnamefont {T.}~\bibnamefont
  {Kontos}},\ }\bibfield  {title} {\enquote {\bibinfo {title} {Coupling a
  quantum dot, fermionic leads, and a microwave cavity on a chip},}\ }\href
  {\doibase 10.1103/PhysRevLett.107.256804} {\bibfield  {journal} {\bibinfo
  {journal} {Phys. Rev. Lett.}\ }\textbf {\bibinfo {volume} {107}},\ \bibinfo
  {pages} {256804} (\bibinfo {year} {2011})}\BibitemShut {NoStop}%
\bibitem [{\citenamefont {Schir\'o}\ and\ \citenamefont
  {Le~Hur}(2014)}]{SchiroLeHurPRB14}%
  \BibitemOpen
  \bibfield  {author} {\bibinfo {author} {\bibfnamefont {M.}~\bibnamefont
  {Schir\'o}}\ and\ \bibinfo {author} {\bibfnamefont {K.}~\bibnamefont
  {Le~Hur}},\ }\bibfield  {title} {\enquote {\bibinfo {title} {Tunable hybrid
  quantum electrodynamics from nonlinear electron transport},}\ }\href@noop {}
  {\bibfield  {journal} {\bibinfo  {journal} {Phys. Rev. B}\ }\textbf {\bibinfo
  {volume} {89}},\ \bibinfo {pages} {195127} (\bibinfo {year}
  {2014})}\BibitemShut {NoStop}%
\bibitem [{\citenamefont {Bruhat}\ \emph {et~al.}(2016)\citenamefont {Bruhat},
  \citenamefont {Viennot}, \citenamefont {Dartiailh}, \citenamefont
  {Desjardins}, \citenamefont {Kontos},\ and\ \citenamefont
  {Cottet}}]{BruhatEtAlPRX16}%
  \BibitemOpen
  \bibfield  {author} {\bibinfo {author} {\bibfnamefont {L.~E.}\ \bibnamefont
  {Bruhat}}, \bibinfo {author} {\bibfnamefont {J.~J.}\ \bibnamefont {Viennot}},
  \bibinfo {author} {\bibfnamefont {M.~C.}\ \bibnamefont {Dartiailh}}, \bibinfo
  {author} {\bibfnamefont {M.~M.}\ \bibnamefont {Desjardins}}, \bibinfo
  {author} {\bibfnamefont {T.}~\bibnamefont {Kontos}}, \ and\ \bibinfo {author}
  {\bibfnamefont {A.}~\bibnamefont {Cottet}},\ }\bibfield  {title} {\enquote
  {\bibinfo {title} {Cavity photons as a probe for charge relaxation resistance
  and photon emission in a quantum dot coupled to normal and superconducting
  continua},}\ }\href@noop {} {\bibfield  {journal} {\bibinfo  {journal} {Phys.
  Rev. X}\ }\textbf {\bibinfo {volume} {6}},\ \bibinfo {pages} {021014}
  (\bibinfo {year} {2016})}\BibitemShut {NoStop}%
\bibitem [{\citenamefont {Cottet}\ \emph {et~al.}(2017)\citenamefont {Cottet},
  \citenamefont {Dartiailh}, \citenamefont {Desjardins}, \citenamefont
  {Cubaynes}, \citenamefont {Contamin}, \citenamefont {Delbecq}, \citenamefont
  {Viennot}, \citenamefont {Bruhat}, \citenamefont {Douçot},\ and\
  \citenamefont {Kontos}}]{CottetEtAlReview17}%
  \BibitemOpen
  \bibfield  {author} {\bibinfo {author} {\bibfnamefont {A.}~\bibnamefont
  {Cottet}}, \bibinfo {author} {\bibfnamefont {M.~C.}\ \bibnamefont
  {Dartiailh}}, \bibinfo {author} {\bibfnamefont {M.~M.}\ \bibnamefont
  {Desjardins}}, \bibinfo {author} {\bibfnamefont {T.}~\bibnamefont
  {Cubaynes}}, \bibinfo {author} {\bibfnamefont {L.~C.}\ \bibnamefont
  {Contamin}}, \bibinfo {author} {\bibfnamefont {M.}~\bibnamefont {Delbecq}},
  \bibinfo {author} {\bibfnamefont {J.~J.}\ \bibnamefont {Viennot}}, \bibinfo
  {author} {\bibfnamefont {L.~E.}\ \bibnamefont {Bruhat}}, \bibinfo {author}
  {\bibfnamefont {B.}~\bibnamefont {Douçot}}, \ and\ \bibinfo {author}
  {\bibfnamefont {T.}~\bibnamefont {Kontos}},\ }\bibfield  {title} {\enquote
  {\bibinfo {title} {Cavity qed with hybrid nanocircuits: from atomic-like
  physics to condensed matter phenomena},}\ }\href@noop {} {\bibfield
  {journal} {\bibinfo  {journal} {Journal of Physics: Condensed Matter}\
  }\textbf {\bibinfo {volume} {29}},\ \bibinfo {pages} {433002} (\bibinfo
  {year} {2017})}\BibitemShut {NoStop}%
\bibitem [{\citenamefont {Liu}\ \emph {et~al.}(2011)\citenamefont {Liu},
  \citenamefont {Li}, \citenamefont {Huang}, \citenamefont {Li}, \citenamefont
  {Guo}, \citenamefont {Laine}, \citenamefont {Breuer},\ and\ \citenamefont
  {Piilo}}]{LiuEtAlNatPhys11}%
  \BibitemOpen
  \bibfield  {author} {\bibinfo {author} {\bibfnamefont {B.-H.}\ \bibnamefont
  {Liu}}, \bibinfo {author} {\bibfnamefont {L.}~\bibnamefont {Li}}, \bibinfo
  {author} {\bibfnamefont {Y.-F.}\ \bibnamefont {Huang}}, \bibinfo {author}
  {\bibfnamefont {C.-F.}\ \bibnamefont {Li}}, \bibinfo {author} {\bibfnamefont
  {G.-C.}\ \bibnamefont {Guo}}, \bibinfo {author} {\bibfnamefont {E.-M.}\
  \bibnamefont {Laine}}, \bibinfo {author} {\bibfnamefont {H.-P.}\ \bibnamefont
  {Breuer}}, \ and\ \bibinfo {author} {\bibfnamefont {J.}~\bibnamefont
  {Piilo}},\ }\bibfield  {title} {\enquote {\bibinfo {title} {Experimental
  control of the transition from markovian to non-markovian dynamics of open
  quantum systems},}\ }\href {http://dx.doi.org/10.1038/nphys2085} {\bibfield
  {journal} {\bibinfo  {journal} {Nat Phys}\ }\textbf {\bibinfo {volume} {7}},\
  \bibinfo {pages} {931--934} (\bibinfo {year} {2011})}\BibitemShut {NoStop}%
\bibitem [{\citenamefont {Ángel Rivas}, \citenamefont {Huelga},\ and\
  \citenamefont {Plenio}(2014)}]{RivasHuelgaPlenioRepProgPhys14}%
  \BibitemOpen
  \bibfield  {author} {\bibinfo {author} {\bibnamefont {Ángel Rivas}},
  \bibinfo {author} {\bibfnamefont {S.~F.}\ \bibnamefont {Huelga}}, \ and\
  \bibinfo {author} {\bibfnamefont {M.~B.}\ \bibnamefont {Plenio}},\ }\bibfield
   {title} {\enquote {\bibinfo {title} {Quantum non-markovianity:
  characterization, quantification and detection},}\ }\href
  {http://stacks.iop.org/0034-4885/77/i=9/a=094001} {\bibfield  {journal}
  {\bibinfo  {journal} {Reports on Progress in Physics}\ }\textbf {\bibinfo
  {volume} {77}},\ \bibinfo {pages} {094001} (\bibinfo {year}
  {2014})}\BibitemShut {NoStop}%
\bibitem [{\citenamefont {Breuer}\ \emph {et~al.}(2016)\citenamefont {Breuer},
  \citenamefont {Laine}, \citenamefont {Piilo},\ and\ \citenamefont
  {Vacchini}}]{BreuerEtalRMP16}%
  \BibitemOpen
  \bibfield  {author} {\bibinfo {author} {\bibfnamefont {H.-P.}\ \bibnamefont
  {Breuer}}, \bibinfo {author} {\bibfnamefont {E.-M.}\ \bibnamefont {Laine}},
  \bibinfo {author} {\bibfnamefont {J.}~\bibnamefont {Piilo}}, \ and\ \bibinfo
  {author} {\bibfnamefont {B.}~\bibnamefont {Vacchini}},\ }\bibfield  {title}
  {\enquote {\bibinfo {title} {\textit{Colloquium} : Non-markovian dynamics in
  open quantum systems},}\ }\href {\doibase 10.1103/RevModPhys.88.021002}
  {\bibfield  {journal} {\bibinfo  {journal} {Rev. Mod. Phys.}\ }\textbf
  {\bibinfo {volume} {88}},\ \bibinfo {pages} {021002} (\bibinfo {year}
  {2016})}\BibitemShut {NoStop}%
\bibitem [{\citenamefont {Schir{\'o}}\ and\ \citenamefont
  {Fabrizio}(2009)}]{schiroFabrizioPRB2009}%
  \BibitemOpen
  \bibfield  {author} {\bibinfo {author} {\bibfnamefont {M.}~\bibnamefont
  {Schir{\'o}}}\ and\ \bibinfo {author} {\bibfnamefont {M.}~\bibnamefont
  {Fabrizio}},\ }\bibfield  {title} {\enquote {\bibinfo {title} {Real-time
  diagrammatic monte carlo for nonequilibrium quantum transport},}\ }\href@noop
  {} {\bibfield  {journal} {\bibinfo  {journal} {Physical Review B}\ }\textbf
  {\bibinfo {volume} {79}},\ \bibinfo {pages} {153302} (\bibinfo {year}
  {2009})}\BibitemShut {NoStop}%
\bibitem [{\citenamefont {Bickers}(1987)}]{Bickers1987}%
  \BibitemOpen
  \bibfield  {author} {\bibinfo {author} {\bibfnamefont {N.~E.}\ \bibnamefont
  {Bickers}},\ }\bibfield  {title} {\enquote {\bibinfo {title} {{Review of
  techniques in the large-N expansion for dilute magnetic alloys}},}\ }\href
  {\doibase 10.1103/RevModPhys.59.845} {\bibfield  {journal} {\bibinfo
  {journal} {Reviews of Modern Physics}\ }\textbf {\bibinfo {volume} {59}},\
  \bibinfo {pages} {845--939} (\bibinfo {year} {1987})}\BibitemShut {NoStop}%
\bibitem [{\citenamefont {Nordlander}\ \emph {et~al.}(1999)\citenamefont
  {Nordlander}, \citenamefont {Pustilnik}, \citenamefont {Meir}, \citenamefont
  {Wingreen},\ and\ \citenamefont {Langreth}}]{Nordlander1999}%
  \BibitemOpen
  \bibfield  {author} {\bibinfo {author} {\bibfnamefont {P.}~\bibnamefont
  {Nordlander}}, \bibinfo {author} {\bibfnamefont {M.}~\bibnamefont
  {Pustilnik}}, \bibinfo {author} {\bibfnamefont {Y.}~\bibnamefont {Meir}},
  \bibinfo {author} {\bibfnamefont {N.~S.}\ \bibnamefont {Wingreen}}, \ and\
  \bibinfo {author} {\bibfnamefont {D.~C.}\ \bibnamefont {Langreth}},\
  }\bibfield  {title} {\enquote {\bibinfo {title} {{How Long Does It Take for
  the Kondo Effect to Develop?}}}\ }\href {\doibase 10.1103/PhysRevLett.83.808}
  {\bibfield  {journal} {\bibinfo  {journal} {Physical Review Letters}\
  }\textbf {\bibinfo {volume} {83}},\ \bibinfo {pages} {808--811} (\bibinfo
  {year} {1999})},\ \Eprint {http://arxiv.org/abs/9903240v1} {arXiv:9903240v1
  [arXiv:cond-mat]} \BibitemShut {NoStop}%
\bibitem [{\citenamefont {Eckstein}\ and\ \citenamefont
  {Werner}(2010)}]{Eckstein_NCA_PRB10}%
  \BibitemOpen
  \bibfield  {author} {\bibinfo {author} {\bibfnamefont {M.}~\bibnamefont
  {Eckstein}}\ and\ \bibinfo {author} {\bibfnamefont {P.}~\bibnamefont
  {Werner}},\ }\bibfield  {title} {\enquote {\bibinfo {title} {Nonequilibrium
  dynamical mean-field calculations based on the noncrossing approximation and
  its generalizations},}\ }\href@noop {} {\bibfield  {journal} {\bibinfo
  {journal} {Phys. Rev. B}\ }\textbf {\bibinfo {volume} {82}},\ \bibinfo
  {pages} {115115} (\bibinfo {year} {2010})}\BibitemShut {NoStop}%
\bibitem [{\citenamefont {R\"uegg}\ \emph {et~al.}(2013)\citenamefont
  {R\"uegg}, \citenamefont {Gull}, \citenamefont {Fiete},\ and\ \citenamefont
  {Millis}}]{rueggMillisPRB2013}%
  \BibitemOpen
  \bibfield  {author} {\bibinfo {author} {\bibfnamefont {A.}~\bibnamefont
  {R\"uegg}}, \bibinfo {author} {\bibfnamefont {E.}~\bibnamefont {Gull}},
  \bibinfo {author} {\bibfnamefont {G.~A.}\ \bibnamefont {Fiete}}, \ and\
  \bibinfo {author} {\bibfnamefont {A.~J.}\ \bibnamefont {Millis}},\ }\bibfield
   {title} {\enquote {\bibinfo {title} {Sum rule violation in self-consistent
  hybridization expansions},}\ }\href {\doibase 10.1103/PhysRevB.87.075124}
  {\bibfield  {journal} {\bibinfo  {journal} {Phys. Rev. B}\ }\textbf {\bibinfo
  {volume} {87}},\ \bibinfo {pages} {075124} (\bibinfo {year}
  {2013})}\BibitemShut {NoStop}%
\bibitem [{\citenamefont {Strand}, \citenamefont {Eckstein},\ and\
  \citenamefont {Werner}(2015)}]{strandWernerPRX2015}%
  \BibitemOpen
  \bibfield  {author} {\bibinfo {author} {\bibfnamefont {H.~U.~R.}\
  \bibnamefont {Strand}}, \bibinfo {author} {\bibfnamefont {M.}~\bibnamefont
  {Eckstein}}, \ and\ \bibinfo {author} {\bibfnamefont {P.}~\bibnamefont
  {Werner}},\ }\bibfield  {title} {\enquote {\bibinfo {title} {Nonequilibrium
  dynamical mean-field theory for bosonic lattice models},}\ }\href {\doibase
  10.1103/PhysRevX.5.011038} {\bibfield  {journal} {\bibinfo  {journal} {Phys.
  Rev. X}\ }\textbf {\bibinfo {volume} {5}},\ \bibinfo {pages} {011038}
  (\bibinfo {year} {2015})}\BibitemShut {NoStop}%
\bibitem [{\citenamefont {Peronaci}, \citenamefont {Schir\'o},\ and\
  \citenamefont {Parcollet}(2018)}]{peronaciPRL2018}%
  \BibitemOpen
  \bibfield  {author} {\bibinfo {author} {\bibfnamefont {F.}~\bibnamefont
  {Peronaci}}, \bibinfo {author} {\bibfnamefont {M.}~\bibnamefont {Schir\'o}},
  \ and\ \bibinfo {author} {\bibfnamefont {O.}~\bibnamefont {Parcollet}},\
  }\bibfield  {title} {\enquote {\bibinfo {title} {Resonant thermalization of
  periodically driven strongly correlated electrons},}\ }\href {\doibase
  10.1103/PhysRevLett.120.197601} {\bibfield  {journal} {\bibinfo  {journal}
  {Phys. Rev. Lett.}\ }\textbf {\bibinfo {volume} {120}},\ \bibinfo {pages}
  {197601} (\bibinfo {year} {2018})}\BibitemShut {NoStop}%
\bibitem [{\citenamefont {Breuer}\ and\ \citenamefont
  {Petruccione}(2007)}]{breuerPetruccione2010}%
  \BibitemOpen
  \bibfield  {author} {\bibinfo {author} {\bibfnamefont {H.}~\bibnamefont
  {Breuer}}\ and\ \bibinfo {author} {\bibfnamefont {F.}~\bibnamefont
  {Petruccione}},\ }\href {https://books.google.co.uk/books?id=DkcJPwAACAAJ}
  {\emph {\bibinfo {title} {The Theory of Open Quantum Systems}}}\ (\bibinfo
  {publisher} {OUP Oxford},\ \bibinfo {year} {2007})\BibitemShut {NoStop}%
\bibitem [{\citenamefont {Aoki}\ \emph {et~al.}(2014)\citenamefont {Aoki},
  \citenamefont {Tsuji}, \citenamefont {Eckstein}, \citenamefont {Kollar},
  \citenamefont {Oka},\ and\ \citenamefont {Werner}}]{aokiWernerRMP2014}%
  \BibitemOpen
  \bibfield  {author} {\bibinfo {author} {\bibfnamefont {H.}~\bibnamefont
  {Aoki}}, \bibinfo {author} {\bibfnamefont {N.}~\bibnamefont {Tsuji}},
  \bibinfo {author} {\bibfnamefont {M.}~\bibnamefont {Eckstein}}, \bibinfo
  {author} {\bibfnamefont {M.}~\bibnamefont {Kollar}}, \bibinfo {author}
  {\bibfnamefont {T.}~\bibnamefont {Oka}}, \ and\ \bibinfo {author}
  {\bibfnamefont {P.}~\bibnamefont {Werner}},\ }\bibfield  {title} {\enquote
  {\bibinfo {title} {Nonequilibrium dynamical mean-field theory and its
  applications},}\ }\href {\doibase 10.1103/RevModPhys.86.779} {\bibfield
  {journal} {\bibinfo  {journal} {Rev. Mod. Phys.}\ }\textbf {\bibinfo {volume}
  {86}},\ \bibinfo {pages} {779--837} (\bibinfo {year} {2014})}\BibitemShut
  {NoStop}%
\bibitem [{\citenamefont {Carmichael}(1999)}]{carmichaelStatistical1999}%
  \BibitemOpen
  \bibfield  {author} {\bibinfo {author} {\bibfnamefont {H.}~\bibnamefont
  {Carmichael}},\ }\href {https://books.google.fr/books?id=ocgRgM-yJacC} {\emph
  {\bibinfo {title} {Statistical Methods in Quantum Optics 1: Master Equations
  and Fokker-Planck Equations}}},\ Physics and Astronomy Online Library\
  (\bibinfo  {publisher} {Springer},\ \bibinfo {year} {1999})\BibitemShut
  {NoStop}%
\bibitem [{\citenamefont {Schwinger}(1961)}]{schwinger1961}%
  \BibitemOpen
  \bibfield  {author} {\bibinfo {author} {\bibfnamefont {J.}~\bibnamefont
  {Schwinger}},\ }\bibfield  {title} {\enquote {\bibinfo {title} {Brownian
  motion of a quantum oscillator},}\ }\href {\doibase 10.1063/1.1703727}
  {\bibfield  {journal} {\bibinfo  {journal} {Journal of Mathematical Physics}\
  }\textbf {\bibinfo {volume} {2}},\ \bibinfo {pages} {407--432} (\bibinfo
  {year} {1961})},\ \Eprint
  {http://arxiv.org/abs/https://doi.org/10.1063/1.1703727}
  {https://doi.org/10.1063/1.1703727} \BibitemShut {NoStop}%
\bibitem [{\citenamefont {Keldysh}(1964)}]{keldysh1964diagram}%
  \BibitemOpen
  \bibfield  {author} {\bibinfo {author} {\bibfnamefont {L.~V.}\ \bibnamefont
  {Keldysh}},\ }\bibfield  {title} {\enquote {\bibinfo {title} {Diagram
  technique for nonequilibrium processes},}\ }\href@noop {} {\bibfield
  {journal} {\bibinfo  {journal} {Zh. Eksp. Teor. Fiz.}\ }\textbf {\bibinfo
  {volume} {47}},\ \bibinfo {pages} {1018} (\bibinfo {year}
  {1964})}\BibitemShut {NoStop}%
\bibitem [{\citenamefont
  {Danielewicz}(1984{\natexlab{a}})}]{DANIELEWICZ1984305}%
  \BibitemOpen
  \bibfield  {author} {\bibinfo {author} {\bibfnamefont {P.}~\bibnamefont
  {Danielewicz}},\ }\bibfield  {title} {\enquote {\bibinfo {title} {Quantum
  theory of nonequilibrium processes ii. application to nuclear collisions},}\
  }\href {\doibase https://doi.org/10.1016/0003-4916(84)90093-9} {\bibfield
  {journal} {\bibinfo  {journal} {Annals of Physics}\ }\textbf {\bibinfo
  {volume} {152}},\ \bibinfo {pages} {305 -- 326} (\bibinfo {year}
  {1984}{\natexlab{a}})}\BibitemShut {NoStop}%
\bibitem [{\citenamefont
  {Danielewicz}(1984{\natexlab{b}})}]{DANIELEWICZ1984239}%
  \BibitemOpen
  \bibfield  {author} {\bibinfo {author} {\bibfnamefont {P.}~\bibnamefont
  {Danielewicz}},\ }\bibfield  {title} {\enquote {\bibinfo {title} {Quantum
  theory of nonequilibrium processes, i},}\ }\href {\doibase
  https://doi.org/10.1016/0003-4916(84)90092-7} {\bibfield  {journal} {\bibinfo
   {journal} {Annals of Physics}\ }\textbf {\bibinfo {volume} {152}},\ \bibinfo
  {pages} {239 -- 304} (\bibinfo {year} {1984}{\natexlab{b}})}\BibitemShut
  {NoStop}%
\bibitem [{\citenamefont {Wagner}(1991)}]{wagnerPRB1991}%
  \BibitemOpen
  \bibfield  {author} {\bibinfo {author} {\bibfnamefont {M.}~\bibnamefont
  {Wagner}},\ }\bibfield  {title} {\enquote {\bibinfo {title} {Expansions of
  nonequilibrium green's functions},}\ }\href {\doibase
  10.1103/PhysRevB.44.6104} {\bibfield  {journal} {\bibinfo  {journal} {Phys.
  Rev. B}\ }\textbf {\bibinfo {volume} {44}},\ \bibinfo {pages} {6104--6117}
  (\bibinfo {year} {1991})}\BibitemShut {NoStop}%
\bibitem [{\citenamefont {Schir\'o}(2010)}]{schiroPRB2009}%
  \BibitemOpen
  \bibfield  {author} {\bibinfo {author} {\bibfnamefont {M.}~\bibnamefont
  {Schir\'o}},\ }\bibfield  {title} {\enquote {\bibinfo {title} {Real-time
  dynamics in quantum impurity models with diagrammatic monte carlo},}\ }\href
  {\doibase 10.1103/PhysRevB.81.085126} {\bibfield  {journal} {\bibinfo
  {journal} {Phys. Rev. B}\ }\textbf {\bibinfo {volume} {81}},\ \bibinfo
  {pages} {085126} (\bibinfo {year} {2010})}\BibitemShut {NoStop}%
\bibitem [{\citenamefont {Gardiner}\ and\ \citenamefont
  {Collett}(1985)}]{gardinerCollettPRA1985}%
  \BibitemOpen
  \bibfield  {author} {\bibinfo {author} {\bibfnamefont {C.~W.}\ \bibnamefont
  {Gardiner}}\ and\ \bibinfo {author} {\bibfnamefont {M.~J.}\ \bibnamefont
  {Collett}},\ }\bibfield  {title} {\enquote {\bibinfo {title} {Input and
  output in damped quantum systems: Quantum stochastic differential equations
  and the master equation},}\ }\href {\doibase 10.1103/PhysRevA.31.3761}
  {\bibfield  {journal} {\bibinfo  {journal} {Phys. Rev. A}\ }\textbf {\bibinfo
  {volume} {31}},\ \bibinfo {pages} {3761--3774} (\bibinfo {year}
  {1985})}\BibitemShut {NoStop}%
\bibitem [{\citenamefont {Werner}\ \emph {et~al.}(2006)\citenamefont {Werner},
  \citenamefont {Comanac}, \citenamefont {de' Medici}, \citenamefont {Troyer},\
  and\ \citenamefont {Millis}}]{wernerMillisPRL2006}%
  \BibitemOpen
  \bibfield  {author} {\bibinfo {author} {\bibfnamefont {P.}~\bibnamefont
  {Werner}}, \bibinfo {author} {\bibfnamefont {A.}~\bibnamefont {Comanac}},
  \bibinfo {author} {\bibfnamefont {L.}~\bibnamefont {de' Medici}}, \bibinfo
  {author} {\bibfnamefont {M.}~\bibnamefont {Troyer}}, \ and\ \bibinfo {author}
  {\bibfnamefont {A.~J.}\ \bibnamefont {Millis}},\ }\bibfield  {title}
  {\enquote {\bibinfo {title} {Continuous-time solver for quantum impurity
  models},}\ }\href {\doibase 10.1103/PhysRevLett.97.076405} {\bibfield
  {journal} {\bibinfo  {journal} {Phys. Rev. Lett.}\ }\textbf {\bibinfo
  {volume} {97}},\ \bibinfo {pages} {076405} (\bibinfo {year}
  {2006})}\BibitemShut {NoStop}%
\bibitem [{\citenamefont {M{\"{u}}ller}\ and\ \citenamefont
  {Stace}(2017)}]{Muller2017}%
  \BibitemOpen
  \bibfield  {author} {\bibinfo {author} {\bibfnamefont {C.}~\bibnamefont
  {M{\"{u}}ller}}\ and\ \bibinfo {author} {\bibfnamefont {T.~M.}\ \bibnamefont
  {Stace}},\ }\bibfield  {title} {\enquote {\bibinfo {title} {{Deriving
  Lindblad master equations with Keldysh diagrams: Correlated gain and loss in
  higher order perturbation theory}},}\ }\href {\doibase
  10.1103/PhysRevA.95.013847} {\bibfield  {journal} {\bibinfo  {journal}
  {Physical Review A}\ }\textbf {\bibinfo {volume} {95}},\ \bibinfo {pages}
  {1--24} (\bibinfo {year} {2017})}\BibitemShut {NoStop}%
\bibitem [{\citenamefont {Reimer}\ and\ \citenamefont
  {Wegewijs}(2018)}]{Reimer2018}%
  \BibitemOpen
  \bibfield  {author} {\bibinfo {author} {\bibfnamefont {V.}~\bibnamefont
  {Reimer}}\ and\ \bibinfo {author} {\bibfnamefont {M.~R.}\ \bibnamefont
  {Wegewijs}},\ }\bibfield  {title} {\enquote {\bibinfo {title}
  {{Density-operator evolution: Complete positivity and the Keldysh real-time
  expansion}},}\ }\href {http://arxiv.org/abs/1808.09395} {\ ,\ \bibinfo
  {pages} {1--45} (\bibinfo {year} {2018})},\ \Eprint
  {http://arxiv.org/abs/1808.09395} {arXiv:1808.09395} \BibitemShut {NoStop}%
\bibitem [{\citenamefont {Gole\ifmmode~\check{z}\else \v{z}\fi{}},
  \citenamefont {Eckstein},\ and\ \citenamefont
  {Werner}(2015)}]{GolezEtAlPRB15}%
  \BibitemOpen
  \bibfield  {author} {\bibinfo {author} {\bibfnamefont {D.}~\bibnamefont
  {Gole\ifmmode~\check{z}\else \v{z}\fi{}}}, \bibinfo {author} {\bibfnamefont
  {M.}~\bibnamefont {Eckstein}}, \ and\ \bibinfo {author} {\bibfnamefont
  {P.}~\bibnamefont {Werner}},\ }\bibfield  {title} {\enquote {\bibinfo {title}
  {Dynamics of screening in photodoped mott insulators},}\ }\href {\doibase
  10.1103/PhysRevB.92.195123} {\bibfield  {journal} {\bibinfo  {journal} {Phys.
  Rev. B}\ }\textbf {\bibinfo {volume} {92}},\ \bibinfo {pages} {195123}
  (\bibinfo {year} {2015})}\BibitemShut {NoStop}%
\bibitem [{\citenamefont {Chen}\ \emph {et~al.}(2016)\citenamefont {Chen},
  \citenamefont {Cohen}, \citenamefont {Millis},\ and\ \citenamefont
  {Reichman}}]{ThetaEtAlPRB16}%
  \BibitemOpen
  \bibfield  {author} {\bibinfo {author} {\bibfnamefont {H.-T.}\ \bibnamefont
  {Chen}}, \bibinfo {author} {\bibfnamefont {G.}~\bibnamefont {Cohen}},
  \bibinfo {author} {\bibfnamefont {A.~J.}\ \bibnamefont {Millis}}, \ and\
  \bibinfo {author} {\bibfnamefont {D.~R.}\ \bibnamefont {Reichman}},\
  }\bibfield  {title} {\enquote {\bibinfo {title} {Anderson-holstein model in
  two flavors of the noncrossing approximation},}\ }\href {\doibase
  10.1103/PhysRevB.93.174309} {\bibfield  {journal} {\bibinfo  {journal} {Phys.
  Rev. B}\ }\textbf {\bibinfo {volume} {93}},\ \bibinfo {pages} {174309}
  (\bibinfo {year} {2016})}\BibitemShut {NoStop}%
\end{thebibliography}
\end{document}